\documentclass[useAMS,usenatbib]{mn2e}

\usepackage{times}
\usepackage{amssymb, amsmath}
\usepackage{graphicx,rotating,amssymb,supertabular,longtable,lscape}
\usepackage{epsfig}
\usepackage[dvipdfm,colorlinks=true,linkcolor=blue,citecolor=blue,urlcolor=blue]{hyperref} 
\def\apj{\mbox{ApJ}}
\def\apjl{\mbox{ApJL}}
\def\apjs{\mbox{ApJS}}

\def\mnras{\mbox{MNRAS}}
\def\aj{\mbox{AJ}}
\def\araa{\mbox{ARA\&A}}
\def\pasp{\mbox{PASP}}
\def\nat{\mbox{Nature}}
\def\aap{\mbox{A\&A}}

\title[AGN dusty tori as clumpy two-phase medium]{Three-dimensional
radiative transfer modeling of AGN dusty tori as a clumpy two-phase
medium}

\author[M. Stalevski et al.]{Marko Stalevski$^{1,2,3}$\thanks{E-mail: mstalevski@aob.rs (MS)}, 
Jacopo Fritz$^{2}$, Maarten Baes$^{2}$, Theodoros Nakos$^{2}$ 
\newauthor and Luka \v{C}. Popovi\'{c}$^{1,3}$
\\
$^{1}$Astronomical Observatory, Volgina 7, 11060 Belgrade, Serbia\\
$^{2}$Sterrenkundig Observatorium, Universiteit Gent, Krijgslaan 281-S9, Gent, 9000, Belgium\\
$^{3}$Isaac Newton Institute of Chile, Yugoslavia Branch}

\begin{document}

\date{\today}

\pagerange{\pageref{firstpage}--\pageref{lastpage}} \pubyear{2011}

\maketitle

\label{firstpage}

\begin{abstract}
We investigate the emission of active galactic nuclei (AGN) dusty
tori in the infrared domain. Following theoretical predictions 
coming from hydrodynamical simulations, we model the dusty torus as a
3D two-phase medium with high-density clumps and low-density medium
filling the space between the clumps. Spectral energy distributions
(SED) and images of the torus at different wavelengths are obtained using
3D Monte Carlo radiative transfer code SKIRT. Our approach of
generating clumpy structure allows us to model tori with single
clumps, complex structures of merged clumps or interconnected
sponge-like structure. A corresponding set of clumps-only models and
models with smooth dust distribution is calculated for comparison. We
found that dust distribution, optical depth, clump size and their
actual arrangement in the innermost region, all have an impact on the
shape of near- and mid-infrared SED. The $10$ $\mu$m silicate feature
can be suppressed for some parameters, but models with smooth dust
distribution are also able to produce a wide range of the silicate
feature strength. Finally, we find that having the dust
distributed in a two-phase medium, might offer a natural solution to
the lack of emission in the near-infrared, compared to observed data,
which affects clumpy models currently available in the literature.
\end{abstract}


\section{Introduction}

According to the unification model, the different appearance of type
1 and type 2 active galactic nuclei (AGN) is only a matter of
orientation \citep{antonucci93, urrypad95}. This hypothesis
postulates a supermassive black hole and its accretion disc as the
central engine. The accretion disc is the source of the strong X-ray
emission and UV/optical continuum. Superimposed on the continuum are
the broad emission lines, coming from gaseous clouds moving at high
velocities, the so-called broad-line region (BLR). This central
region is surrounded by a geometrically and optically thick toroidal
structure of dust and gas with an equatorial visual optical depth
much larger than unity. Viewed edge-on, this dusty torus blocks the
radiation coming from the accretion disc and BLR. In this case an
UV/optical bump and broad emission lines are absent and the object
appears as a type 2 AGN. In the case when the line of sight
does not cross the dusty torus, both accretion disc and BLR
are exposed, giving rise to a strong UV/optical continuum and broad
emission lines, and the object is classified as a type 1 active
galaxy. Observed polarized nuclear emission in type 2 sources,
scattered by electrons and tenuous dust \citep{antonmiller85,
pieranton94, pack97}, supports the unification model. It proves that
an active galactic nucleus is present even though no direct emission
from accretion disc is observed. The toroidal geometry also explains
several other observables such as, the presence of ionizing cones
\citep{pogg88,pogg89,tadtsv89} and high hydrogen column densities in
the X-ray, usually associated with type 2 sources
\citep[e.g.][]{shi06}.

The dusty torus is expected to absorb the incoming accretion disc
radiation and re-emit it in the infrared domain. Thus, further
support for the existence of a dusty torus comes from the observed
mid- to far-infrared bump and silicate feature at $\sim 10$ $\mu$m in
the spectral energy distribution (SED) of AGNs. In type 1 sources,
hot dust in the inner region can be seen directly and the feature is
expected to be detected in emission. Recent mid-infrared observations
obtained with the \textit{Spitzer} satellite confirm the silicate
emission
feature in AGNs \citep{Sie05,Hao05}. In type 2 objects, the
dust feature is observed in absorption \citep[e.g.,][]{Jaf04} due to
obscuration by the cold dust. Additional evidence of dusty tori comes
from the interferometric observations. Using speckle interferometry,
the nucleus of NGC\,1068 was successfully resolved in the $K$-band
\citep{Wit98} and in the $H$-band \citep{Wei04}. This resolved core
is interpreted as dust close to the sublimation radius.
\citet{trist07} reported VLTI interferometric observations with
strong
evidence of a parsec scale dust structure in the Circinus galaxy.
\citet{kish11} reported indication of a partial resolution of the
dust
sublimation region in several type 1 AGNs using the Keck
interferometer.

In order to prevent the dust grains from being destroyed by the hot
surrounding gas, \citet{krolikbegel88} suggested that the dust in the
torus is organized in a large number of optically thick clumps.
However, due to the difficulties in handling clumpy media and lack of
computational power, early work was conducted by adopting a smooth
dust distribution. Several authors explored different radial and
vertical density profiles \citep{pierkrolik92, pierkrolik93,
granatodanese94, efstathiourowan95, vanbemmeldullemond03,
schartmann05, fritz06}. The first effort of developing the formalism
for the treatment of clumpy media was undertaken by
\citeauthor{nenkova02} (\citeyear{nenkova02}, \citeyear{nenkova08a},
\citeyear{nenkova08b}). They utilized a 1D radiative transfer code to
compute the SED of a single irradiated clump. In a second step a
statistical generalization is made to assemble the SED of the torus.
They claim that only clumpy tori are able to reproduce the observed
properties of the silicate feature. However, \citet{dullemond05}
modeled the torus as a whole, using a 2D radiative transfer
calculations. They adopted a geometry with axial symmetry and modeled
clumps in the form of rings around the polar axis. They made a direct
comparison of models with clumps and corresponding smooth dust
distributions and concluded that there is no evidence for a
systematic suppression of the silicate emission feature in the clumpy
models. \citet{honig06}, with an upgrade of their model in
\citet{honig10}, adopted a similar method as \citet{nenkova02}, but
they employed a 2D radiative transfer code and took into account
various illumination patterns of clumps. They reported that a
suppression of the silicate emission feature strongly depends on the
random distribution and density of the clumps in the innermost
region. More recently \citet{schartmann08} presented a 3D
radiative transfer models of clumpy tori. Their findings are in
agreement with those by \citet{honig06}. On the other hand, using
the models developed by \citet{fritz06}, \citet{feltre11} found that
a
smooth distribution of dust is also capable of reproducing the
observed variety of the silicate feature strength. A further study of
the silicate feature and its properties, such as the appearance in
emission in some type 2 objects and the apparent shifting toward long
wavelengths in some objects, is presented in \citet{nikutta09}.

A problem which the obscuring torus hypothesis faced from the
beginning was formation of the dynamically stable structure and
maintenance of the required scale-height. \citet{krolikbegel88}
presented a scenario according to which the scale-height is
maintained through regular elastic collisions between the clumps
\citep[see also][]{beckdusch04}. In the case of a continuous dust
distribution, \citet{pierkrolik92}, followed by \citet{krolik07},
suggested that radiation pressure within the torus may be enough to
support the structure. \citet{wadanorman02} \citep[with model update
in][]{wada09} performed a 3D hydrodynamical simulations, taking into
account self-gravity of the gas, radiative cooling and heating due to
supernovae. They found that such a turbulent medium would produce a
multiphase filamentary (sponge-like) structure, rather then isolated
clumps. A scenario where the effects of stellar feedback from a
nuclear cluster play a major role is discussed in
\citet{schartmann09}.

In this paper we present our modeling of 3D AGN dusty tori. We model
the torus as the whole, using the 3D Monte Carlo radiative transfer
code SKIRT. We take a step further toward a more realistic model by
treating the dusty torus as a two-phase medium, with high density
clumps and low density medium filling the space between them. We
calculated SEDs and images of the torus for a grid of parameters. Our
approach allows us to, for each two-phase model, generate a
clumps-only model (with dust distributed to the clumps exclusively,
without any dust between them) and a smooth model with the same
global physical parameters. Our aims are (a)\ to investigate the
influence of different parameters on model SEDs and their observable
properties, (b)\ to put to a test reports that the observed SEDs in
the mid-infrared domain unambiguously point to a clumpy structure of
dusty tori; if that is indeed the case, a comparison of clumpy and
smooth models should show a systematic difference of their observable
properties, such as the strength of the silicate feature.

The paper is organized as follows. In Section \ref{sec:mod} we give
the description of the radiative transfer code and present the
physical details of our model. In Section \ref{sec:res} we discuss
how different parameters affect modeled SEDs, analyze their
observable properties and compare two-phase, clumps-only and smooth
models. Finally, in Section \ref{sec:conc} we outline our
conclusions.

\section{Model}
\label{sec:mod}

\subsection{The radiative transfer code}

We have used the radiative transfer code SKIRT \citep{baes03, baes11}
for the modelling of AGN dusty tori. SKIRT is a 3D continuum
radiative transfer code based on the Monte Carlo algorithm
\citep{cashwelleverett59, witt77}, initially developed to study the
effect of dust absorption and scattering on the observed kinematics
of dusty galaxies \citep{baesdejonghe01, baesdejonghe02, baes03}. It
has been extended with a module to self-consistently calculate the
dust emission spectrum under the assumption of local thermal
equilibrium -- LTE \citep{baes05a}. This LTE version of SKIRT has
been used to model the dust extinction and emission of galaxies and
circumstellar discs \citep{baes10, delooze10, vidalbaes07}.

Similar to most modern Monte Carlo codes \citep[e.g.][]{gordon01,
wolf03, niccolini03, bianchi08}, the SKIRT code contains many
deterministic elements which makes the Monte Carlo technique orders
of magnitude more efficient than the naive Monte Carlo recipe. These
include the peeling-off technique \citep{yusefzadeh84}, continuous
absorption \citep{lucy99, niccolini03}, forced scattering
\citep{cashwelleverett59, witt77} and smart detectors \citep{baes08}.
For the current simulations, we use the technique of frequency
distribution adjustment presented by \citet{bjorkmanwood01} and
critically discussed by \citet{baes05b}. This technique ensures that
at each moment during the simulation, the wavelength distribution
from the photon packages emitted from the cell are in agreement with
the cell's current temperature. The main advantage of this
technique is that no iteration is necessary.

\subsection{Model description}

\subsubsection{Dust distribution and properties}

\begin{figure}
\centering
\includegraphics[height=0.30\textwidth]{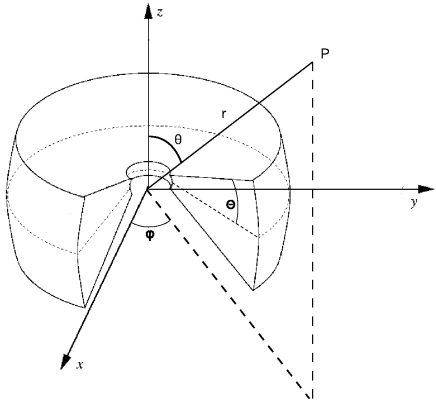}
\caption{Schematic representation of the adopted model geometry and
coordinate system.}
\label{fig:schem}
\end{figure}

We approximate the spatial dust distribution around the primary
continuum source with a conical torus (i.e. a flared disc). Its
characteristics are defined by (a)\ the half opening angle $\Theta$
-- defining also the maximum height extent to which the dust is
distributed --, (b)\ the inner and outer radius, $R_{\text{in}}$ and
$R_{\text{out}}$ respectively, and (c) the parameters describing dust
density distribution, $p$ and $q$ (see below). A schematic
representation of the adopted geometry is given in Fig.
\ref{fig:schem}. For the inner radius of the dusty torus we adopted
the value of $0.5$ pc, at wich the dust grains are heated to the
temperature of $\sim 1180$ K, according to the prescription given by
\citet{barvainis87}:
\begin{equation}\label{eqn:rmin}
R_{\text{in}}\simeq 1.3 \cdot \sqrt{L_{46}^{AGN}}\cdot T_{1500}^{-2.8} \qquad \text{[pc]},
\end{equation}
where $L_{46}^{AGN}$ is the bolometric ultraviolet/optical luminosity
emitted by the central source, expressed in units of $10^{46}$ erg
s$^{-1}$ and $T_{1500}$ is the sublimation temperature of the dust
grains given in units of 1500~K.

We describe the spatial distribution of the dust density with a law
that allows a density gradient along the radial direction and with
polar angle, as the one adopted by \citet{granatodanese94}:
%---------------------------------
\begin{equation}\label{eqn:dens}
\rho\left(r,\theta \right)\propto r^{-p}e^{-q|cos(\theta)|} ,
\end{equation}
%---------------------------------
where $r$ and $\theta$ are coordinates in the adopted polar
coordinate system (see Fig. \ref{fig:schem}).
%---------------------------------
%---------------------------------

The dust mixture consists of separate populations of graphite and
silicate dust grains with a classical MRN size distribution
\citep*{mrn77}:
%---------------------------------
\begin{equation}\label{eqn:grains}
dn(a)=Ca^{-3.5}da ,
\end{equation}
%---------------------------------
where the size of grains, $a$, varies from $0.005$ to $0.25$ $\mu$m
for both graphite and silicate. The normalization factors for size
distribution are $C=10^{-15.13}$ and $C=10^{-15.11}$ for graphite and
silicate, respectively \citep{weingdrain01}. Optical properties are
taken from \citet{laordraine93} and \citet{lidraine01}.

The dust is distributed on a 3D cartesian grid composed of a large
number of cubic cells. The dust density is constant within each cell.
The standard resolution for our simulations is 200 cells along each
axis ($8\times10^{6}$ cells in total). However, to properly sample
the dust properties, an increase in the torus size requires an
increase of the resolution of the computational grid as well. Thus,
to simulate a torus twice the size of the `standard model', one
needs to employ a grid with $400$ cells along each axis, that is,
$64\times10^{6}$ cells in total. 

In the case of a smooth density distribution, the axial symmetry in
our model reduces the 3D radiative transfer computations to a 2D
problem, with a significant gain both on the computational time and
the memory usage. However, the prescription we use for generating
clumpy models demands a 3D cartesian grid. Therefore, such a grid was
used throughout all our simulations, in order to avoid any possible
differences due to the adoption of different grids. To ensure that
the adopted resolution is sufficient to properly sample the dust, for
each simulation we compare the actual and expected values of (a)\ the
face-on and edge-on central surface density and (b)\ the total dust
mass.

The emission for all models was calculated on an equally spaced
logarithmic wavelength grid ranging from $0.001$ to $1000$ $\mu$m. A
finer wavelength sampling was adopted between $5$ and $35$ $\mu$m, in
order to better resolve the shape of $10$ and $18$ $\mu$m silicate
features. Each simulation is calculated using $10^{8}$ photon
packages.

\subsubsection{Spectral energy distribution of the primary source}
\label{sec:accdsk}

\begin{figure}
\centering
\includegraphics[height=0.34\textwidth]{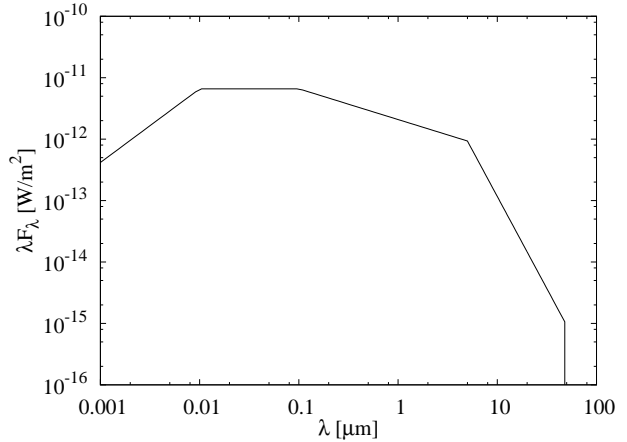}
\caption{The SED of the primary source, i.e. the accretion disc,
which we assume to irradiate as a composition of power laws,
with different indices for different wavelengths ranges (see text for
details).}
\label{fig:source}
\end{figure}

The primary continuum source of dust heating is the intense
UV-optical continuum coming from the accretion disc. 
A very good approximation of its emission is a central, point-like
energy source, emitting isotropically. Its SED is very well
approximated by a composition of power laws with different spectral
indices in different spectral ranges. The adopted values are:
\begin{equation}\label{eqn:source}
\lambda L(\lambda)\propto\left\{
\begin{array}{lrrr}
\lambda^{1.2}  &  \; 0.001 \leq \lambda < \leq 0.01 & [\mu\text{m}]\\
\lambda^{0}    &  \; 0.01  < \lambda \leq 0.1   & [\mu\text{m}]\\
\lambda^{-0.5} &  \; 0.1   < \lambda \leq 5     & [\mu\text{m}]\\
\lambda^{-3}   &  \; 5     < \lambda \leq 50    & [\mu\text{m}]
\end{array}
\right.
\end{equation}
and the resulting SED is plotted on Fig. \ref{fig:source}.  These
values have been quite commonly adopted in the literature, and come
from both observational and theoretical arguments \citep[see
e.g.,][]{schartmann05}. We have anyway verified that changes in the
shape of the primary source SED affect only very marginally the
infrared re-emission. For the bolometric luminosity of the primary
source we adopted the value of $10^{11}$ $L_{\odot}$ \citep[see
e.g.,][]{davislaor11}.

%---------------------------------
%---------------------------------

As mentioned above, an isotropic emission of the central source
is commonly adopted in the literature \citep[e.g.][]{schartmann05,
honig06, nenkova08a}. However, the disk emission is inevitably
anisotropic \citep[see, for example,][and references therein]{kawmor11}.
Therefore, we have performed additional calculations assuming
anisotropic radiation of the central source, in order to investigate
the resulting influence on the model SEDs. Radiation flux ($F$) from
a unit surface area of an optically thick disk toward a unit solid
angle at the polar angle of $\theta$  decreases with an increasing
$\theta$ according to the formula given by \citet{netz87}:
%-----------------
\begin{equation}\label{eqn:anisof}
F\propto \cos\theta(1+2\cos\theta) ,
\end{equation}
%-----------------
where the first term represents the change in the projected surface
area, and the the latter represents the limb darkening effect. For
simplicity, in our calculations we take into account only the first
term. As the accretion disk radiation varies with $\theta$, the dust
sublimation radius also depends on it:
%-----------------
\begin{equation}\label{eqn:anisor}
R_{in}=R_{in,0}(2|\cos\theta|)^{0.5} ,
\end{equation}
%-----------------
where $R_{in,0}$ is the dust sublimation radius estimated assuming
isotropic emission. As a result, the inner edge of the torus is
(a)\ closer to the central source than suggested by the
Eq.~\ref{eqn:rmin} and (b)\ the structure of the edge is concave
\citep{kawamor10,kawmor11}. Also, \citet{kawamor10} found that the
dust sublimation radius can decrease down to $0.1R_{in,0}$, all the
way to the outermost radius of accretion disk. However, due to the
numerical constraints, that is, the minumum size of the dust cell in
the computational grid we are currently limited to, we cannot allow
the dust to be placed all the way to the primary source.
Instead, we are forced to put a limit on the minimum allowed dust
sublimation radius at $0.225$ pc ($0.45R_{in,0}$). We discuss the
influence of the anisotropic central source radiation on the dusty
tori model SEDs in the Sec.~\ref{sec:aniso}; throughout the rest of
the paper, the isotropic emission is assumed.
%--------------------------------------------------------------------
\subsubsection{Two-phase medium: the approach}
\label{sec:smooth_clumpy}
%--------------------------------------------------------------------

%---------------------------------
\begin{figure}
\centering
\includegraphics[height=0.47\textwidth]{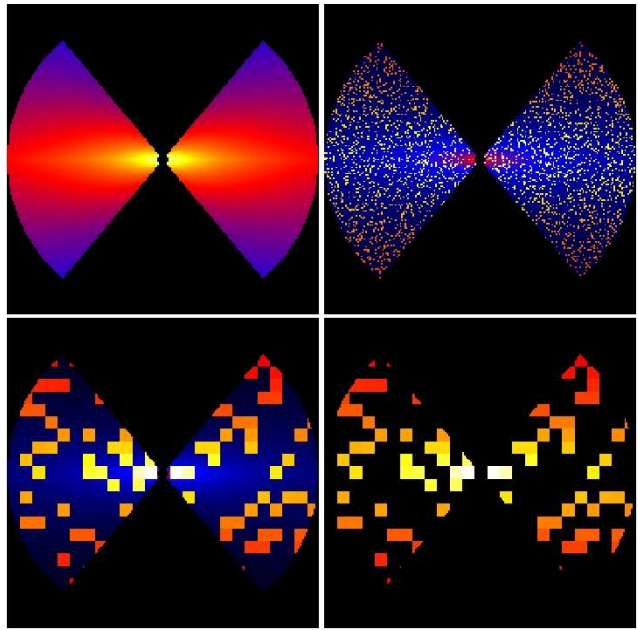}
\caption{Dust density distribution at the meridional plane, given in
logarithmic color scale. Density law parameters are $p=1$ and $q=2$.
The smooth dust distribution is presented in the top left panel.
The top right and bottom left panel present two-phase dust
distribution for two different sizes of clumps: each clump is
composed of one cubic grid cell (top right) and $8\times8\times8$
grid cells (bottom left). In the bottom right panel, a clumps-only
dust distribution is presented. The contrast parameter between high-
and low-density phases in the two-phase and clumps-only models is
$100$ and $10^9$, respectively.}
\label{fig:dens}
\end{figure}
%---------------------------------

Models of emission in which the dust is uniformly, smoothly
distributed within the toroidal region are obtained starting from Eq.
\ref{eqn:dens}. The density gradient is determined by the two
parameters, $p$ and $q$. The total amount of dust is fixed based on
the equatorial optical depth at $9.7$ $\mu$m ($\tau_{9.7}$). While
for the smooth models distributing the dust is straightforward, for
the clumpy model this process is non-trivial. As hydrodynamical
simulations of \citet{wadanorman02} demonstrated, dust in tori is
expected to take the form of a multiphase structure, rather than
isolated clumps. Therefore, we adopted the approach which allows us
to generate such a medium. We start from the corresponding smooth
models, i.e. the one with the same global parameters, and then apply
the algorithm described by \citet{wittgordon96} to generate a
two-phase clumpy medium. According to this algorithm, each individual
cell in the grid is assigned randomly to either a high- or
low-density state by a Monte Carlo process. The medium created in
such a way is statistically homogeneous, but clumpy. The filling
factor determines the statistical frequency of the cells in the
high-density state and can take values between $0$ and $1$. For
example, a filling factor of $0.01$ represents a case of rare, single
high-density clumps in an extended low-density medium. The process
for the clump generation is random with respect to the spatial
coordinates of the individual clumps themselves. Thus, as the filling
factor is allowed to increase, the likelihood that adjoining cells
are occupied by clumps increases as well. This leads to the
appearance of complex structures formed by several merged clumps. For
filling factor values $\gtrsim0.25$, clumps start to form an
interconnected sponge-like structure, with low-density medium filling
the voids. We form larger clumps by forcing high-density state into
several adjoining cells.

To tune the density of the clumps and the inter-clump medium, we use
the `contrast parameter', defined as the ratio of the dust density
in the high- and low-density phase. This parameter can be assigned
any positive value. For example, setting the contrast to one would
result in a smooth model. Setting extremely high value of contrast
($>1000$) effectively puts all the dust into the clumps, without
low-density medium between them. An example of dust density
distributions at the meridional plane for smooth, two-phase and
clumps-only models is given in Fig. \ref{fig:dens}.

%--------------------------------------------------------------------
\subsection{Parameter grid}
%--------------------------------------------------------------------

In this section we present the adopted values of parameters
used to calculate a grid of models for our analysis.

For the inner and outer radius of the torus we adopted the values
$0.5$ and $15$ pc, respectively. The half opening angle of the torus
-- $\Theta$ -- is fixed to $50^\circ$ for all of our model
realizations. All models are calculated at 7 different line-of-sight
inclinations, namely $0^\circ$, $40^\circ$, $50^\circ$, $60^\circ$,
$70^\circ$, $80^\circ$ and $90^\circ$, where $i=0^\circ$ represents a
face-on (type 1) AGN and $i=90^\circ$ an edge-on (type 2) AGN.
Inclinations between $0^\circ$ and $40^\circ$ (dust-free lines of
sight) were omitted since their SED shows no appreciable difference
with respect to the full face-on view. The equatorial optical depth
$\tau_{9.7}$ takes the values $0.1$, $1.0$, $5.0$, $10.0$. We note
here that this is the optical depth of the initial smooth model,
before the dust is redistributed to make the clumpy one (see
Sec.~\ref{sec:smooth_clumpy}). Thus, the exact value along the given
line of sight will vary due to the random distribution of the clumps.
The parameters defining the spatial distribution of the dust density
(Eq.~\ref{eqn:dens}) are $p=0$, $1$ and $q=0$, $2$, $4$, $6$.

Defining the relative clump size, $\sigma$, as the ratio of the outer
radius of the torus over the clump size:
%-----------------
\begin{equation}\label{eqn:size}
\sigma=R_{out}/D_{clump}
\end{equation}
%-----------------
we explored the clumpy models for two different clump sizes, $0.15$
pc and $1.2$ pc, that is, $\sigma = 100$ and $\sigma = 12.5$,
respectively. The number of clumps in the former case is
$9\times10^5$, and each clump occupies one grid cell. In the latter
case there are $\sim 3000$ clumps, each one being composed of
$8\times8\times8$ grid cells. The adopted filling factor values are
$0.15$ in the case $\sigma = 100$, and $0.25$ in the case $\sigma =
12.5$ models. Both values allow us to have single, as well as
clusters of high-density clumps immersed into a low-density medium.
The contrast between high- and low-density phases is fixed at $100$. 

%---------------------------------
\begin{figure}
\centering
\includegraphics[height=0.34\textwidth]{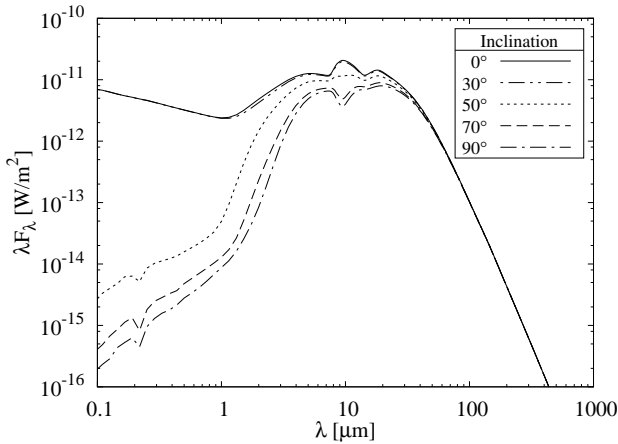}
\caption{Dependence of the modeled SED on the viewing angle:
$i=0^\circ$ (solid line), $i=30^\circ$ (dash-double-dotted),
$i=50^\circ$ (dotted), $i=70^\circ$ (dashed), $i=90^\circ$
(dash-dotted). The first two, almost identical (in fact, they lay
upon each other) are associated with dust-free paths. The inner and
outer radii of the torus are $0.5$ and $15$ pc, respectively, the
half opening angle of the torus, $\Theta$, is $50^\circ$, the optical
depth is $\tau_{9.7}=5$, the parameters of the dust density
distribution are $p=1$ and $q=2$, the clump size is $1.2$ pc (or
$\sigma = 12.5$), the filling factor $0.25$ and the contrast
parameter $100$.}
\label{fig:inc}
\end{figure}
%---------------------------------

We generated three sets of models with the same global physical
parameters: (a)\ models with the dust distributed smoothly,
(b)\ models with the dust as a two-phase medium and (c)\ models with
a contrast parameter set to an extremely high value ($10^9$),
effectively putting all the dust into the high-density clumps. We
will refer to these models as `smooth', `two-phase' and
`clumps-only', respectively. For clumpy models (both two-phase and
clumps-only) we generated another set of models with the same
parameters, but with a different random distribution of clumps. 

For each model we calculated SEDs and images of torus for all
the points in the wavelength grid. Calculated flux is scaled for the
torus distance of 10 Mpc from the observer. The parameter grid is
summarized in Table \ref{tab:par_grid}.

%---------------------------------
%---------------------------------
\begin{table}
\caption{The grid of parameters for which the models have been
computed.}
\centering
\begin{tabular}{l c l}
\hline
Parameter	&  & Adopted values  \\
\hline
L		&  & $10^{11}$ $L_{\odot}$ \\
$R_{in}$	&  & $0.5$ pc \\
$R_{out}$	&  & $15$ pc \\
$\tau_{9.7}$	&  & $0.1$, $1.0$, $5.0$, $10.0$ \\
$p$		&  & $0$, $1$ \\
$q$		&  & $0$, $2$, $4$, $6$ \\
$\Theta$	&  & $50^\circ$ \\
Filling factor	&  & $0.15$, $0.25$ \\
Contrast	&  & $100$ , $10^{9}$\\
Size of clumps	&  & $0.15$ pc, $1.2$ pc \\
Inclination	&  & $0^\circ$, $40^\circ$, $50^\circ$, $60^\circ$,
$70^\circ$, $80^\circ$, $90^\circ$ \\
\hline
\end{tabular}
\label{tab:par_grid}
\end{table}
%---------------------------------
%---------------------------------

%--------------------------------------------------------------------
\section{Results and discussion}
\label{sec:res}
%--------------------------------------------------------------------
In this section we discuss the influence of different parameters on
the general shape of the SEDs of the two-phase models and analyze
their observable properties. The following analysis refers to the
two-phase models with $\sigma = 12.5$. We will address how the
properties of these models compare to properties of models with
$\sigma = 100$, clumps-only, and smooth models in Sections
\ref{sec:smooth} and \ref{sec:clonly}.

%--------------------------------------------------------------------
\subsection{SED dependence on the viewing angle}
\begin{figure*}
\centering
\includegraphics[height=0.49\textwidth]{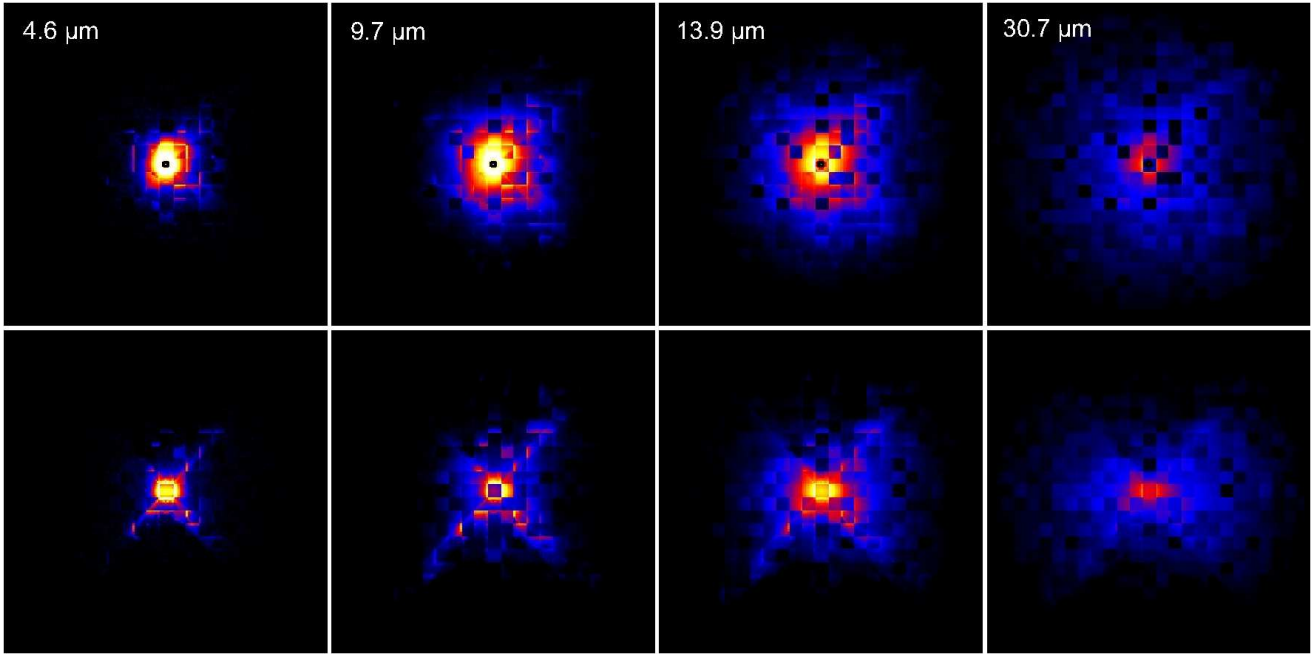}
\caption{Images of the torus at different wavelengths. Top row is
face-on view, bottom row edge-on view. From left to right, panels
represent images at $4.6$, $9.7$, $13.9$, and $30.7$ $\mu$m. Images
are given in logarithmic color scale. The parameters are the same as
in Fig. \ref{fig:inc}. The visible squared structure is due to the
clumps which in the model are in the form of cubes.}
\label{fig:img}
\end{figure*}

\begin{figure*}
\centering
\includegraphics[height=0.34\textwidth]{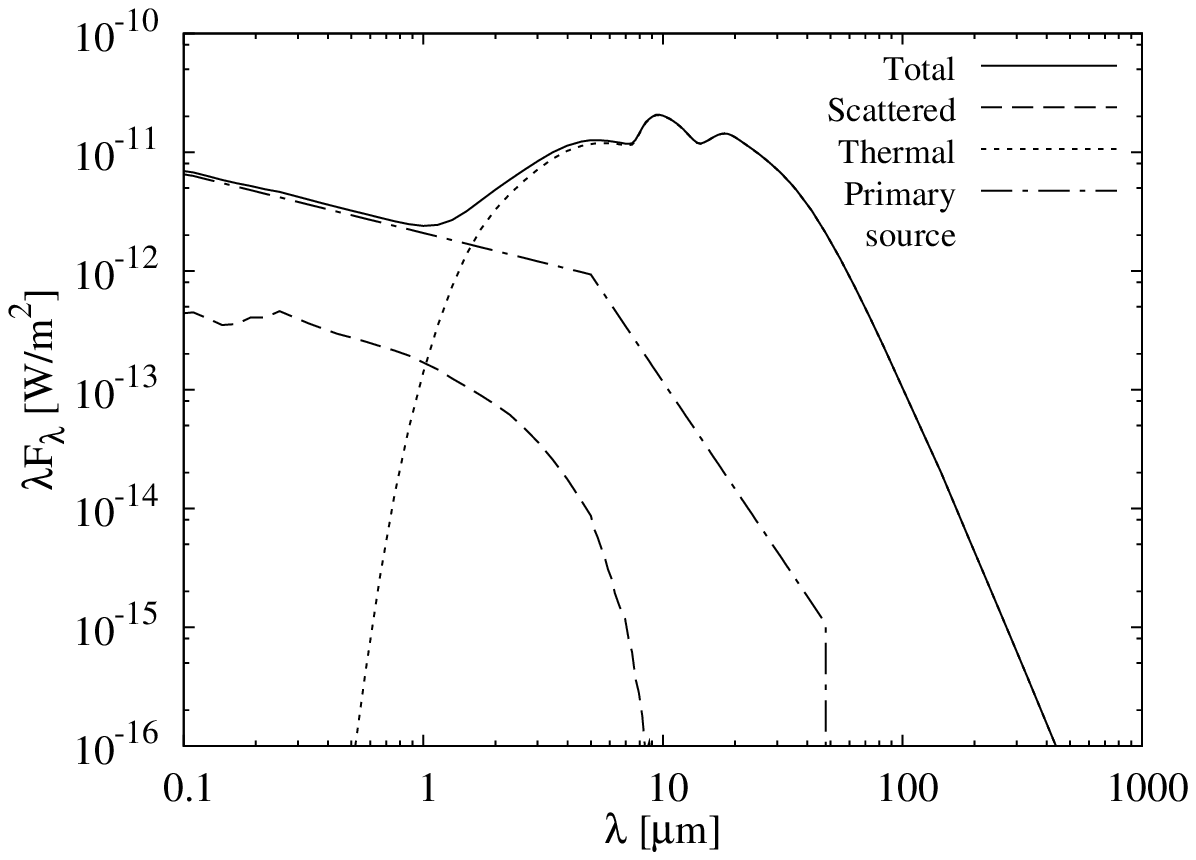}
\includegraphics[height=0.34\textwidth]{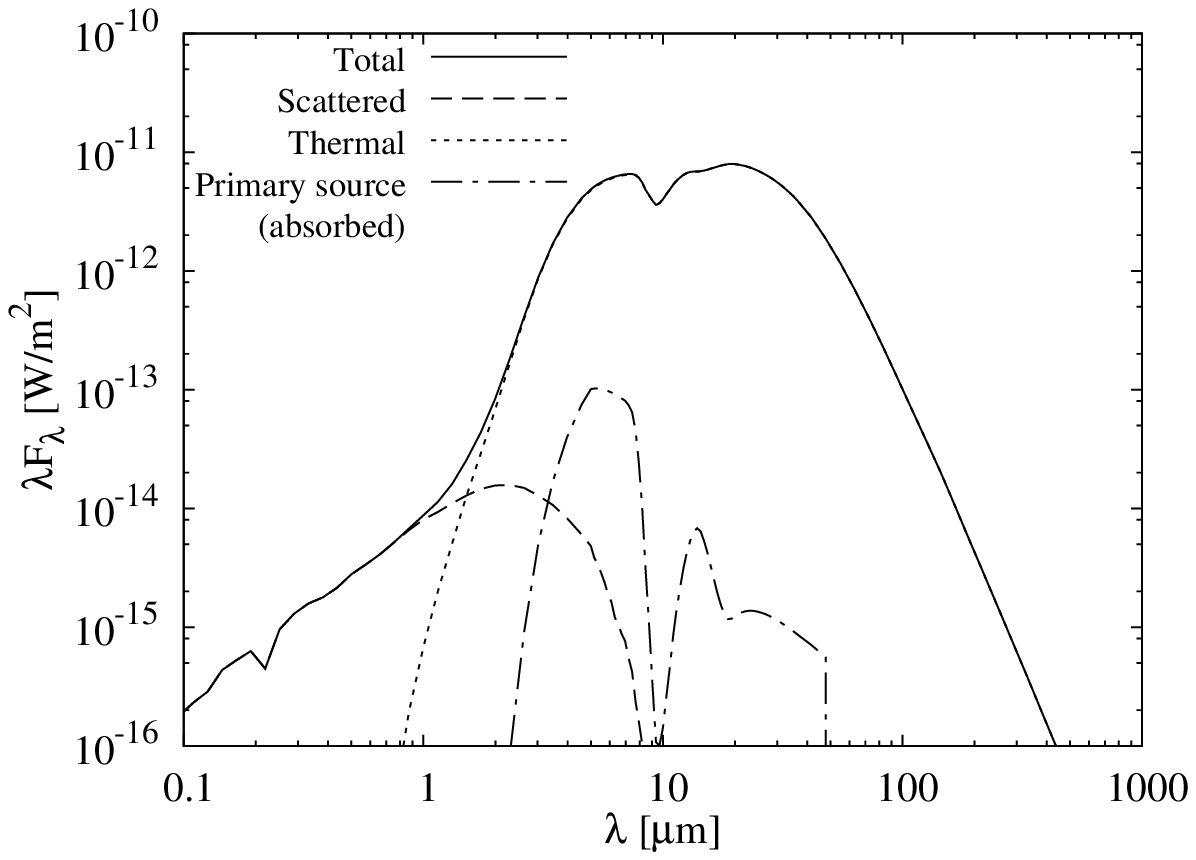}
\caption{The total (solid line), thermal (dotted line), scattered
(dashed line) and primary source (dash-dotted line) emission are
plotted. Left panel: face-on view; right panel: edge-on view. The
parameters are the same as in Fig. \ref{fig:inc}.}
\label{fig:flux}
\end{figure*}

%--------------------------------------------------------------------
As it was demonstrated in early works
\citep[e.g.][]{granatodanese94}, the SED of a dusty torus depends on
the viewing angle. In Fig. \ref{fig:inc} we show the SED dependence
on the inclination of the torus. There is a clear distinction between
cases of dust-free lines of sight ($i=0^\circ$, $30^\circ$) and those
that pass through the torus ($i=50^\circ$, $70^\circ$, $90^\circ$).
For the adopted value of half opening angle ($\Theta=50^\circ$), this
transition occurs at inclination $i=40^\circ$. The most notable
difference is found shortward of $1$ $\mu$m. In the case of dust-free
lines of sight, we directly see the radiation coming from the
accretion disc, while in the case of dust-intercepting paths most of
the radiation is absorbed and scattered at different wavelengths.
This is especially pronounced in those systems where the density
remains constant with polar angle. In the case of a non-constant
density, the transition from face-on to edge-on view is
smoother. An exception is the case of very low optical depths, where
it is possible to directly `see' the central source even when
viewed edge-on. Another important difference between dust-free and
dust-intercepting lines of sight is the $10$ $\mu$m silicate feature,
which is expected to appear in emission in the former case and in the
absorbtion in the latter. The properties of this feature is
discussed in detail in Section \ref{sec:silfeat}. Images of the torus
at different wavelengths, for the type 1 and type 2 lines of sight
are shown in Fig. \ref{fig:img}. From the figure it is clear that, at
shorter wavelengths, it is the radiation from the inner (and hotter)
region that dominates. At longer wavelengths, the emission arises
from the dust placed further away. Thus, the size of torus is
wavelength dependent. In Fig. \ref{fig:flux} we present the total SED
and its thermal and scattered components, along with the primary
source emission, for face-on and edge-on view. As it can be
seen from the figure, the thermal component is dominant in the mid-
and far-infrared part of SED and its shape is similar for both type 1
and type 2 orientations. The shape and amount of scattered component
is quite different; in the edge-on view it determines the total SED
at shorter wavelengths, while in the face-on view it is negligible
compared to the primary source emission.

%---------------------------------
%---------------------------------

%---------------------------------
%---------------------------------

%--------------------------------------------------------------------
\subsection{SED dependence on the filling factor and contrast}
%--------------------------------------------------------------------

\begin{figure}
\centering
\includegraphics[height=0.43\textwidth]{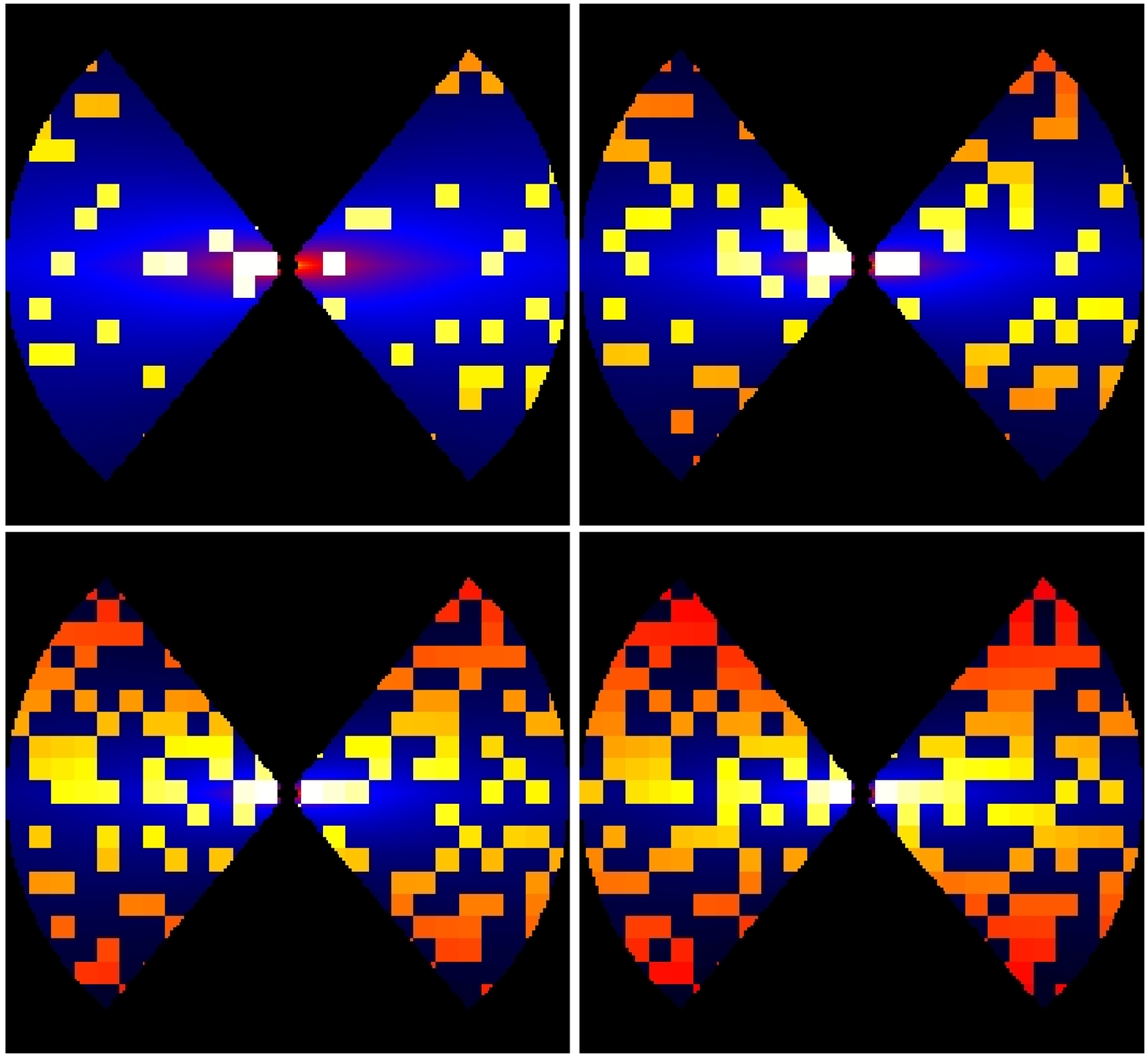}
\caption{Dust density distribution at the meridional plane for
different filling factors, in logarithmic color scale. The filling
factors are: 0.15 (top left panel), 0.25 (top right), 0.35 (bottom
left), and 0.45 (bottom right). All other parameters are the same as
in Fig. \ref{fig:inc}.}
\label{fig:ff_dens}
\end{figure}

\begin{figure*}
\centering
\includegraphics[height=0.34\textwidth]{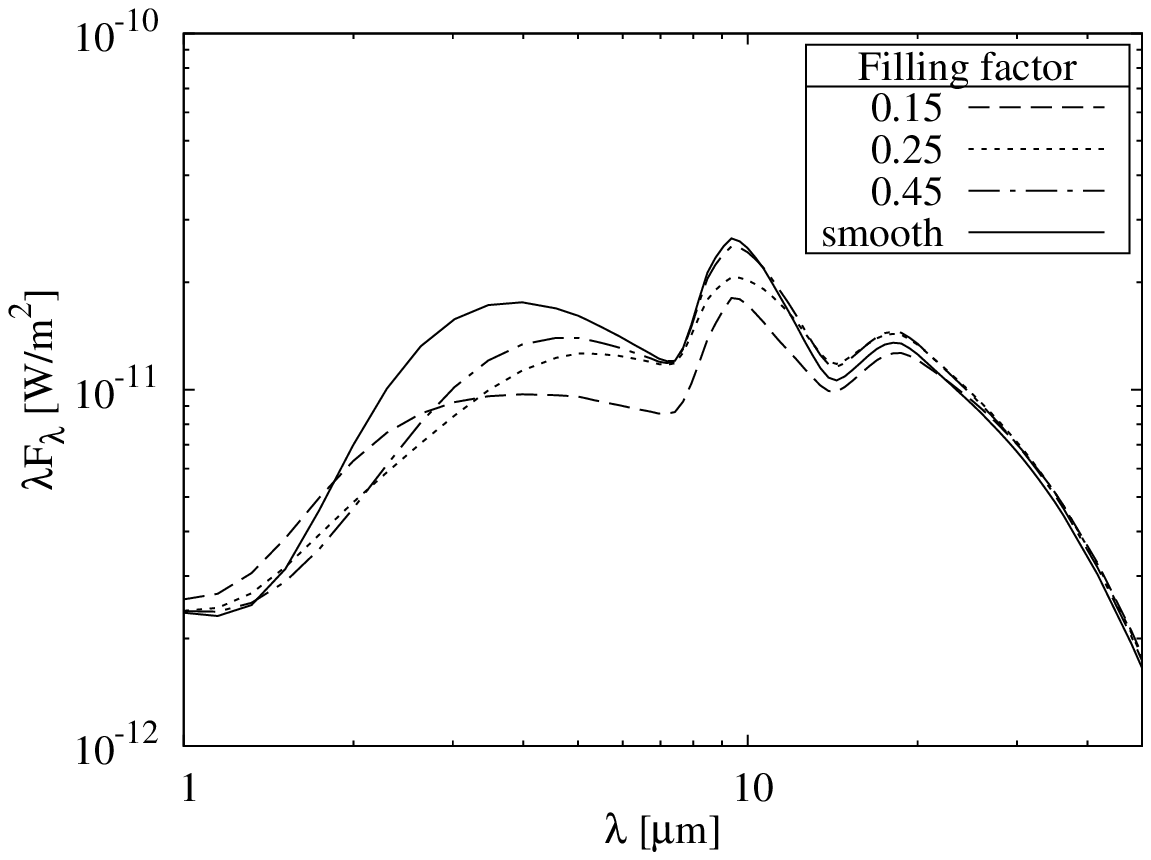}
\includegraphics[height=0.34\textwidth]{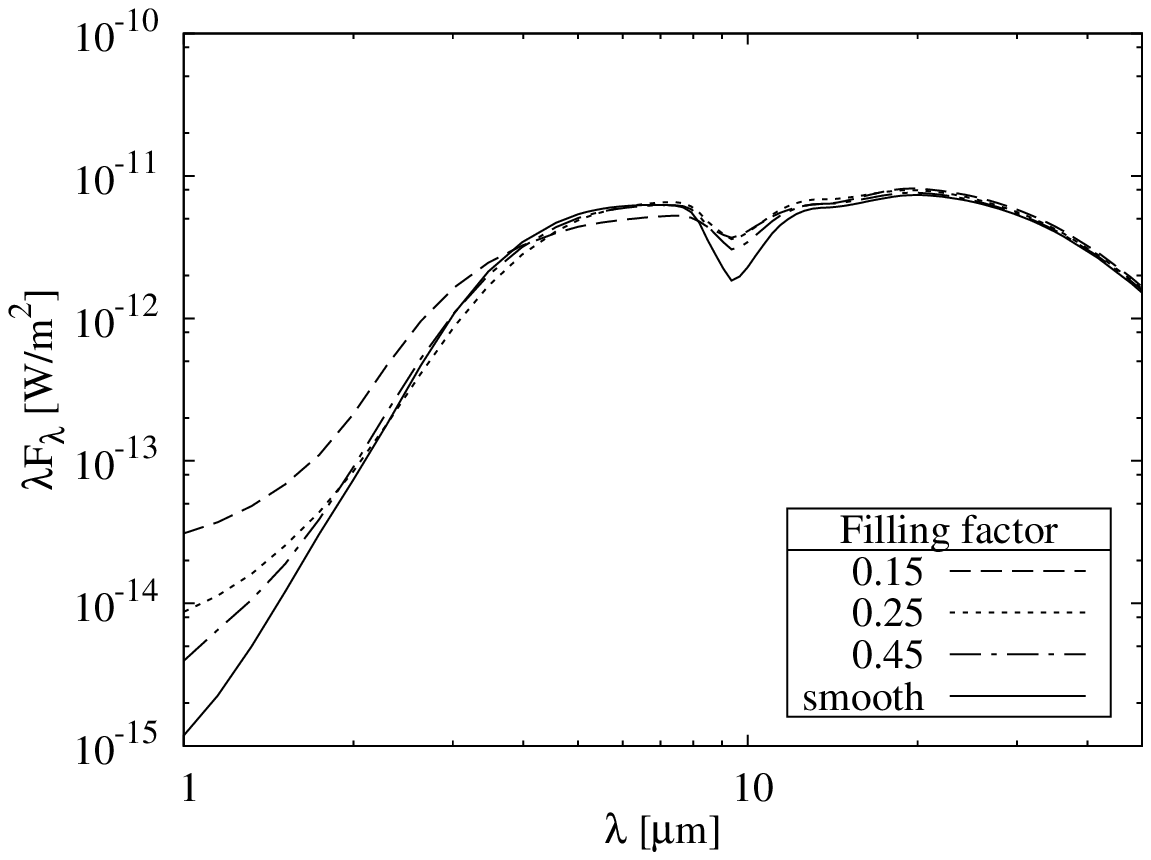}
\caption{Model SEDs for different filling factors: dashed line:
0.15; dotted line: 0.25; dash-dotted line: 0.45. For comparison
purposes,  the SED of a corresponding smooth model is also plotted
(solid line). All other parameters are the same as in Fig.
\ref{fig:inc}. Left panel: face-on view; right panel: edge-on view.}
\label{fig:ff_sed}
\end{figure*}

As described in Section \ref{sec:smooth_clumpy}, the two parameters
that determine the characteristics of the two-phase medium are the
filling factor and the contrast. The filling factor determines the
percentage of grid cells in a high-density state. Models with low
values for the filling factor (e.g. $<0.1$) represent systems with
rare, single high-density clumps in extended low-density medium. As
the filling factor increases, the number of clumps will increase as
well, forming clusters of clumps, or even single, interconnected
sponge-like structure. This is illustrated in Fig.~\ref{fig:ff_dens},
where we show dust density distributions at the meridional plane for
different filling factors. Fig.~\ref{fig:ff_sed} shows SEDs of
models for different filling factors, compared with SED of the
corresponding smooth model. From this Fig. we see that, as the
filling factor increases, the overall mid-infrared emission increases
as well. For filling factor value of $\sim 0.25$, in the face-on
view, the silicate feature is attenuated. As the filling factor
increases, clumps start to form sponge-like structures, more and more
resembling a smooth dust distribution, and the strength of the
silicate feature approaches the strength in the corresponding smooth
model. In the edge-on, a filling factor value of $\sim 0.25$
produces silicate features in weaker absorption than in the
corresponding smooth model. As the filling factor increases, the
strength of the silicate feature approaches the strength of the
feature in the smooth models.

%---------------------------------
%---------------------------------

%---------------------------------
%---------------------------------

%---------------------------------
\begin{figure*}
\centering
\includegraphics[height=0.34\textwidth]{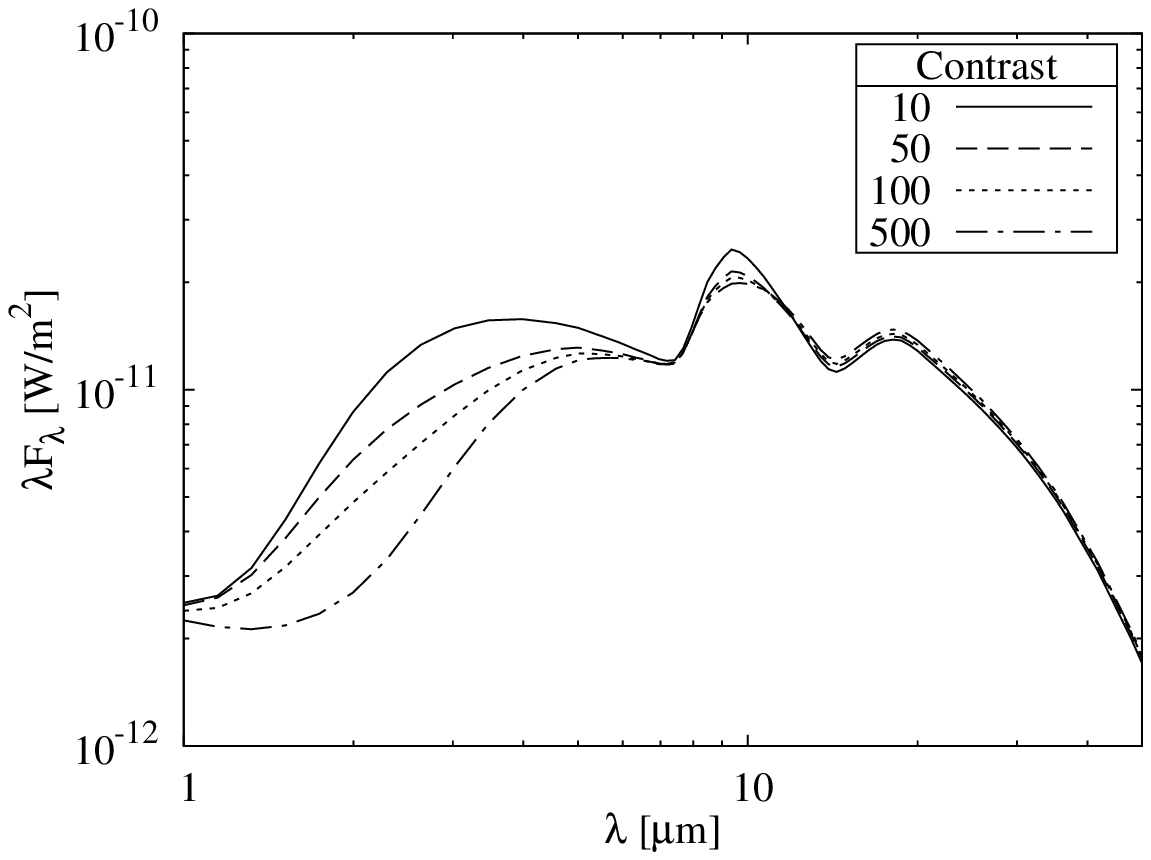}
\includegraphics[height=0.34\textwidth]{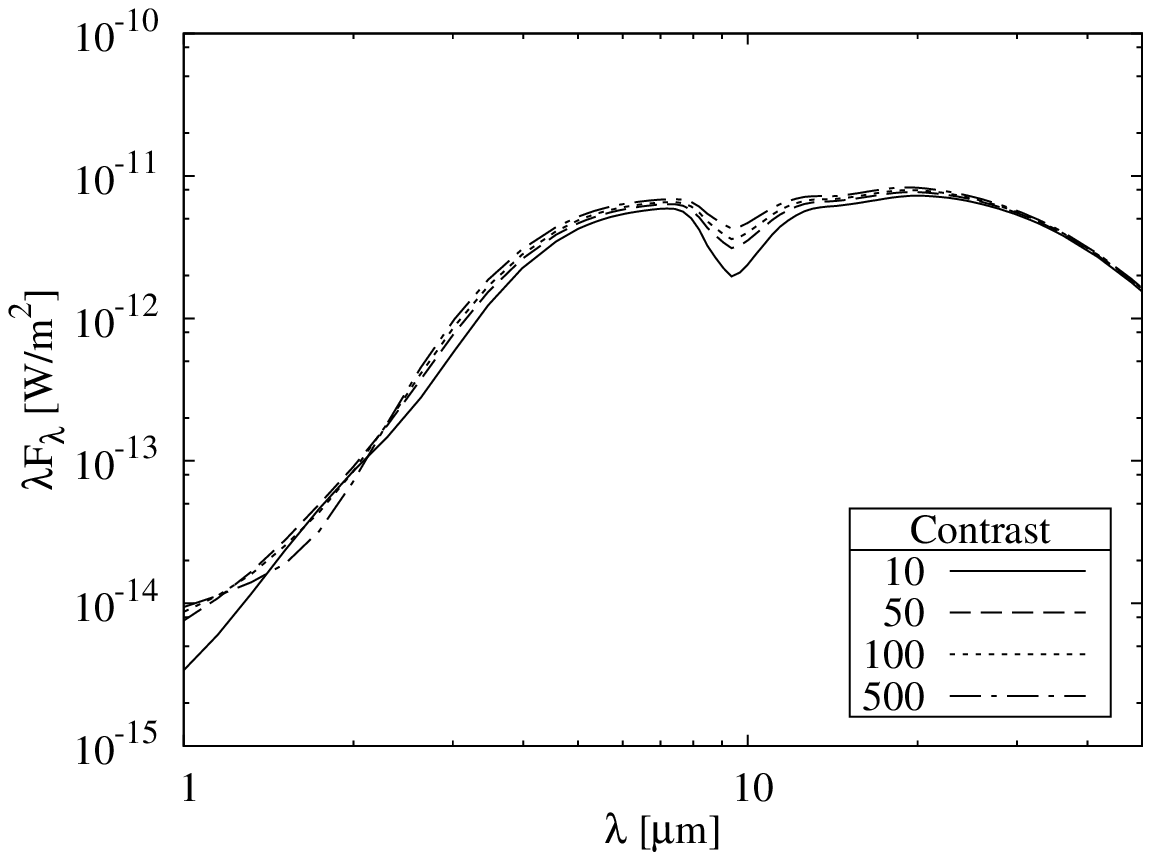}
\caption{Model SEDs for different values of contrast parameter.
Contrast have values of 10 for solid line, 50 for dashed, 100 for
dotted, and 500 for dash-dotted. All other parameters are the same as
in Fig. \ref{fig:inc}. Left panel: face-on view; right panel: edge-on
view.}
\label{fig:contrast}
\end{figure*}
%---------------------------------

The `contrast' parameter sets the density ratio between the high-
and low-density phases. Fig. \ref{fig:contrast} shows the model SED
dependence on this parameter. In the face-on view, for increasing
contrast, both the hot dust emission ($\sim 1 - 6$ $\mu$m) and the
strength of the silicate feature decrease. From the same figure we
also see that, for higher contrast values, the peak of the
silicate feature is slightly shifted toward longer wavelengths. In
the edge-on view, the silicate feature in absorption gets weaker with
increasing contrast.

%--------------------------------------------------------------------
\subsection{SED dependence on the random distribution of clumps}
%--------------------------------------------------------------------

%---------------------------------
\begin{figure*}
\centering
\includegraphics[height=0.23\textwidth]{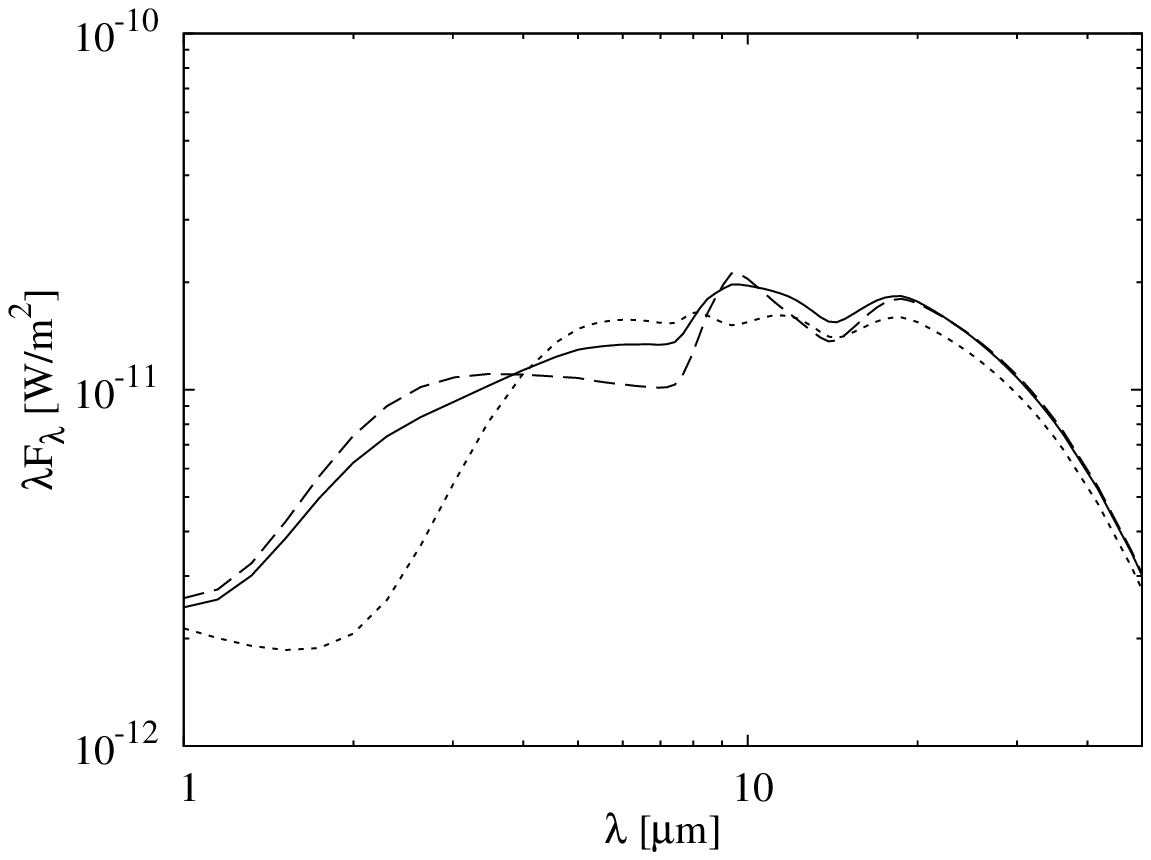}
\includegraphics[height=0.23\textwidth]{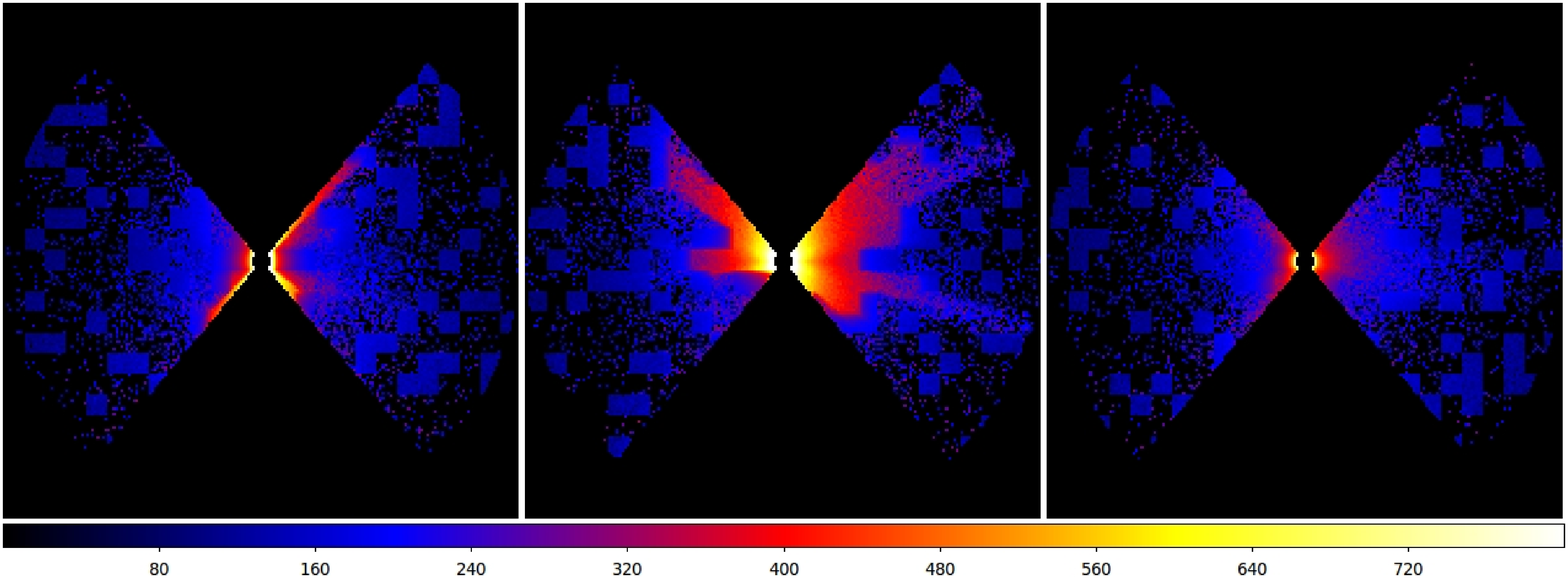}
\caption{Temperature distribution at the meridional plane for the
three different random distributions of clumps (three rightmost
panels) and corresponding SEDs for face-on view (left panel). The
solid line corresponds to the left panel, the dashed to the middle,
and the dotted to the right. The parameters are the same as in
Fig.~\ref{fig:inc}. Temperature images are shown in a logarithmic
color scale, which is for clarity of the images cut off at $800$ K.}
\label{fig:rnd}
\end{figure*}
%---------------------------------

The shape and overall near- and mid-infrared emission strongly depend
on the distribution of dust in the innermost region. Changing the
random arrangement of clumps, along with choosing a particular line
of sight, can affect the resulting SED significantly, as illustrated
in Fig. \ref{fig:rnd}. As described in Section
\ref{sec:smooth_clumpy}, the process for clump generation is random
with respect to the spatial coordinates of the individual clumps. As
a consequence, adjoining cells can be occupied by individual clumps,
forming complex structures of several connected clumps. In models
with a higher concentration of clumps in the innermost region, due to
the shadowing effect, the absorption is increased and silicate
feature is suppressed.

This characteristic imports a degree of degeneracy in the features of
the SEDs, which will be less directly dependent on the physical input
parameters. Even though the spatial position of the clumps is not
related to the physical properties of dusty tori, their
re-arrangement has a clear impact on the infrared emission. It is,
in some way, mimicking a change in the optical depth, which might
appear either to increase or decrease, depending on the clumps
re-arrangement, especially in the innermost regions.

Some random arrangements of the clumps have interesting
repercussions. Because of clumpiness, the difference between the SED
of type 1 and 2 objects is not truly an issue of orientation; it is
rather a matter of probability of directly viewing the main energy
source of the AGN \citep{nenkova08b}. As a result, type 1 sources can
be detected even from what are typically considered as type 2
orientations. Such a scenario provides an explanation for the few
Seyfert galaxies with type 1-like optical spectra whose $0.4-16$
$\mu$m SED resembles that of a type 2 AGN \citep{alher03}.
Conversely, if a clump happens to obscure the central engine from an
observer, that object would be classified as type 2 irrespective of
the viewing angle. In such cases, the clump may move out of the
line-of-sight, creating a clear path to the nucleus and a transition
to a type 1 spectrum. Such transitions between type 1 and type 2 line
spectra have been observed in a few sources \citep[see][and
references therein]{aretx99}.

%--------------------------------------------------------------------
\subsection{Influence of anisotropic primary source radiation on
model SED}
\label{sec:aniso}
\begin{figure}
\centering
\includegraphics[height=0.34\textwidth]{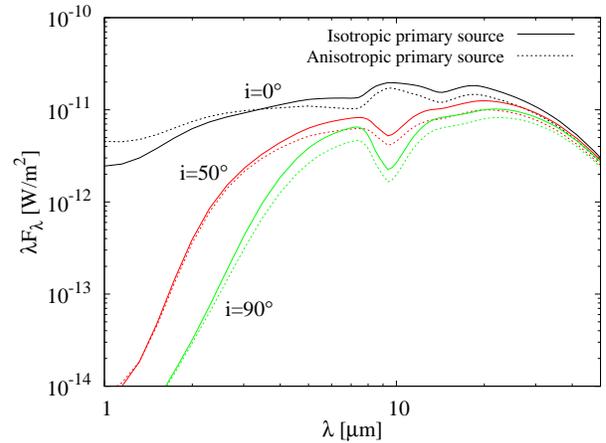}
\caption{Model SEDs assuming isotropic (solid line) and anisotropic
(dotted line) accretion disk radiation. Inclinations are indicated
in the plot. All the torus parameters are the same as in Fig.
\ref{fig:inc}, except for the $q$ parameter which here takes the
value of $0$.}
\label{fig:aninc}
\end{figure}

%--------------------------------------------------------------------
As described in Sec.~\ref{sec:accdsk}, an isotropic source
emission is commonly adopted in the radiative transfer modeling of
dusty tori; however, the accretion disk emission is actually 
anisotropic. In this section, we discuss the influence of
anisotropic source radiation on the model SEDs. Dependence of the
accretion disk radiation on the direction is taken according
to Eq.~\ref{eqn:anisof} and the corresponding change of the dust
sublimation radius according to Eq.~\ref{eqn:anisor}. In
Fig.~\ref{fig:aninc} we present the resulting model SEDs if the
anisotropic radiation of the primary source is assumed (dotted line)
and compare them to the corresponding SEDs obtained in the case of
the isotropic source (solid line) for different inclinations. SEDs
were calculated for the inclinations between $0^\circ$ and $90^\circ$
with the step of $10^\circ$; for the clarity of the Figure, only SEDs
for three inclinations, $0^\circ$, $50^\circ$ and $90^\circ$ are
shown. 

We found that, when anisotropy of the central source is assumed, the
IR SED can indeed change, resulting in a lower emission, though
roughly keeping the same shape. This is a logical consequence coming
from the fact that, for a given bolometric luminosity of the
accretion disc, an anisotropic source whose characteristics are those
as described above, is emitting more power in the dust-free region:
the overall result is a less luminous torus. The excess of emission
shortward of $\sim3\, \mu$m is seen in the dust-free lines of sight,
because, at these wavelengths the primary source contribution is
still significant. Again, as expected from the properties of
radiative transfer \citep{ivezic97}, we found that the shape and the
features of the SED (e.g. the $10\, \mu$m feature) are not affected.
Therefore we conclude that  our analysis in the rest of the paper is
not affected by the isotropic approximation for the central source
radiation.

%--------------------------------------------------------------------
\subsection{The silicate feature strength}
\label{sec:silfeat}
%--------------------------------------------------------------------
As it was mentioned above, an important characteristic in the
infrared part of an AGN SED is the so-called silicate feature. This
silicate feature is caused by Si-O stretching modes, giving rise to
either emission or absorption band, peaking at $\sim10 \, \mu$m. All
of the early models were dealing with the following issue: while they
were properly predicting it in absorption in type 2 objects --in
agreement with what was indeed observed--, observations from that
period were not supporting the models' prediction of a silicate
feature in emission in type 1 AGN. In fact, one of the main issues
driving the development of clumpy models, aimed at addressing this
discrepancy between models and observations. Later observations
performed by \textit{Spitzer} with its infrared spectrometer IRS,
showed that
for a number of type 1 object this feature is indeed observed in
emission, partially solving this issue. Recently, \citet{hony11}
reported the detection of a very strong $10\, \mu$m feature in
emission. On the other hand, \citet{fritz06} showed that smooth
models are also able to properly reproduce the observed emission in
this range. Furthermore, the comparative study performed by
\citet{feltre11} showed that clumpy and smooth dust distributions are
equally able to reproduce both observed broad-band SEDs and
mid-infrared \textit{Spitzer} spectra.

The strength of the $10\, \mu$m feature can be characterized by the
dimensionless parameter $S$, the natural logarithm of the
peak-over-continuum ratio \citep{pierkrolik92, granatodanese94}. The
continuum is defined by a power law connecting the fluxes at $6.8$
and $13.9$ $\mu$m. $S$ assumes positive values for a feature in
emission and negative ones if it is in absorption.

In a face-on view, $S$ takes values in the range $\sim 0.1 - 1$. The
silicate feature is present in a strong emission in the models with
lower optical depths ($\tau_{9.7}=0.1, 1$). Models with an optical
depth of $\tau_{9.7}=5$ are showing a wider range of intensities,
most of them of moderate strength, with a few cases of strong or weak
emission. The strength of the feature in models with high optical
depth ($\tau_{9.7}=10$) is also showing a wide range of intensities,
but with overall lower values, and is significantly attenuated in
some cases.

\begin{figure*}
\centering
\includegraphics[height=0.34\textwidth]{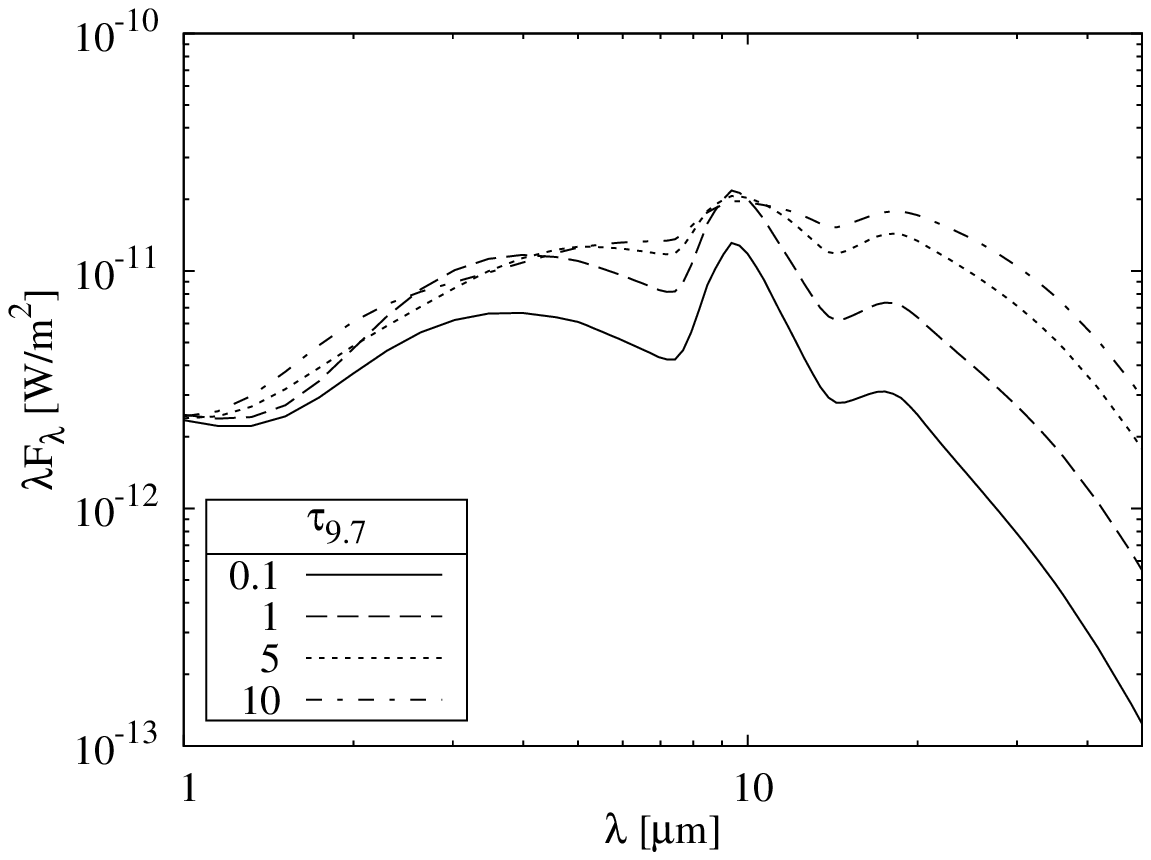}
\includegraphics[height=0.34\textwidth]{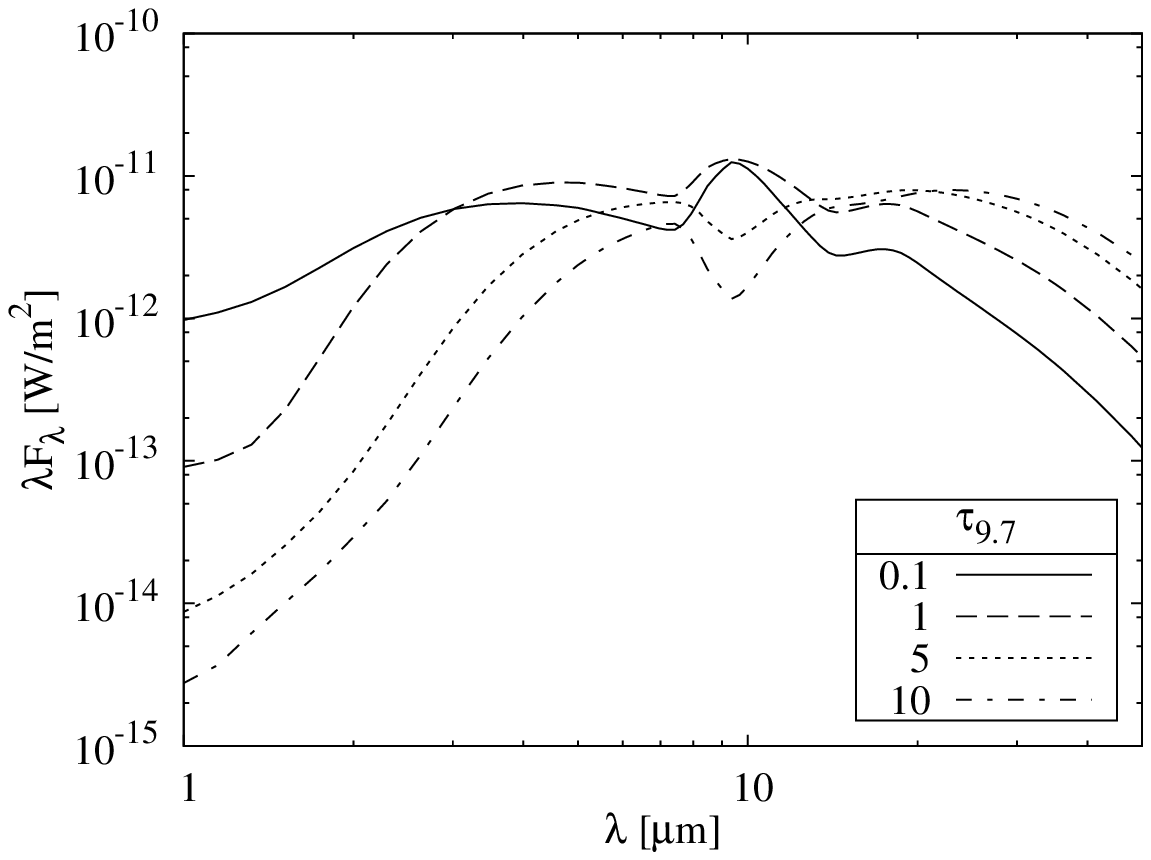}
\caption{Model SEDs for different optical depths. The solid line
represents the case of an optical depth of
$\tau_{9.7}=0.1$, the dashed of $\tau_{9.7}=1.0$, the dotted of
$\tau_{9.7}=5.0$, and the dash-dotted of $\tau_{9.7}=10.0$. All other
parameters are the same as in Fig. \ref{fig:inc}. Left
panel: face-on view; right panel: an edge-on view.}
\label{fig:tau}
\end{figure*}

\begin{figure*}
\centering
\includegraphics[height=0.3\textwidth]{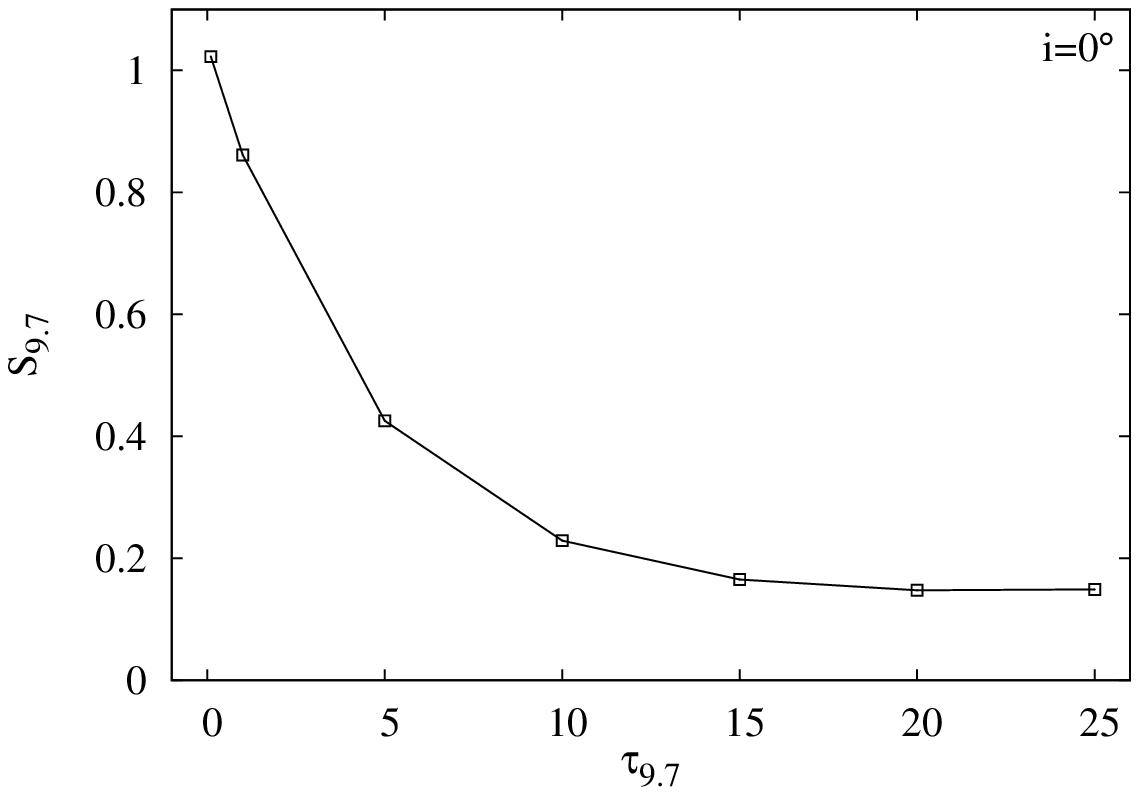}
\includegraphics[height=0.3\textwidth]{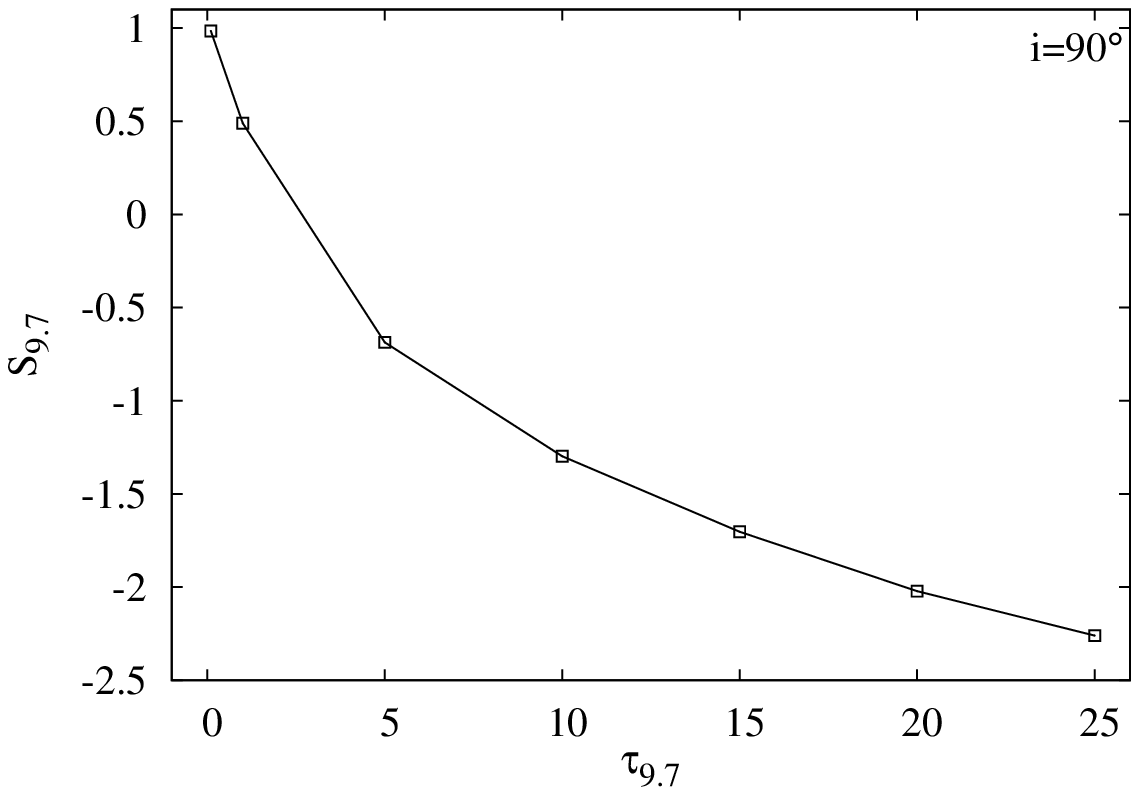}\\
\includegraphics[height=0.3\textwidth]{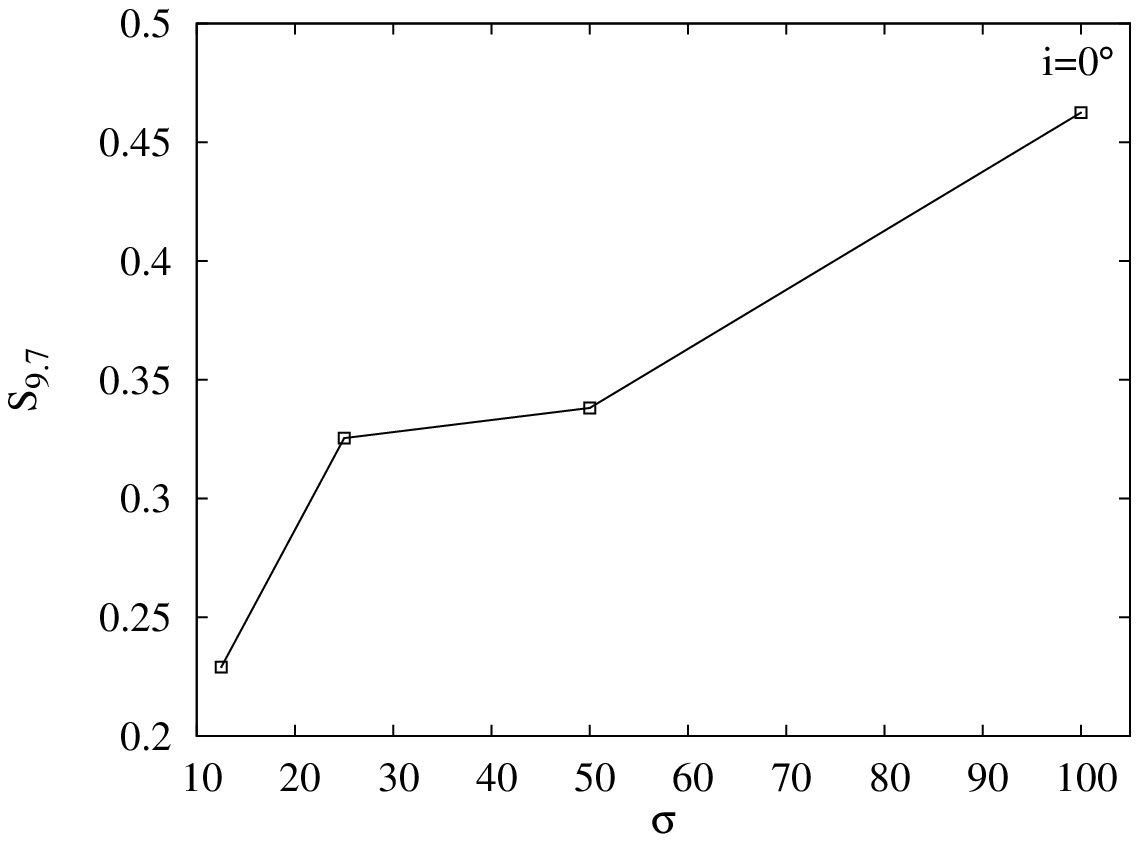}
\includegraphics[height=0.3\textwidth]{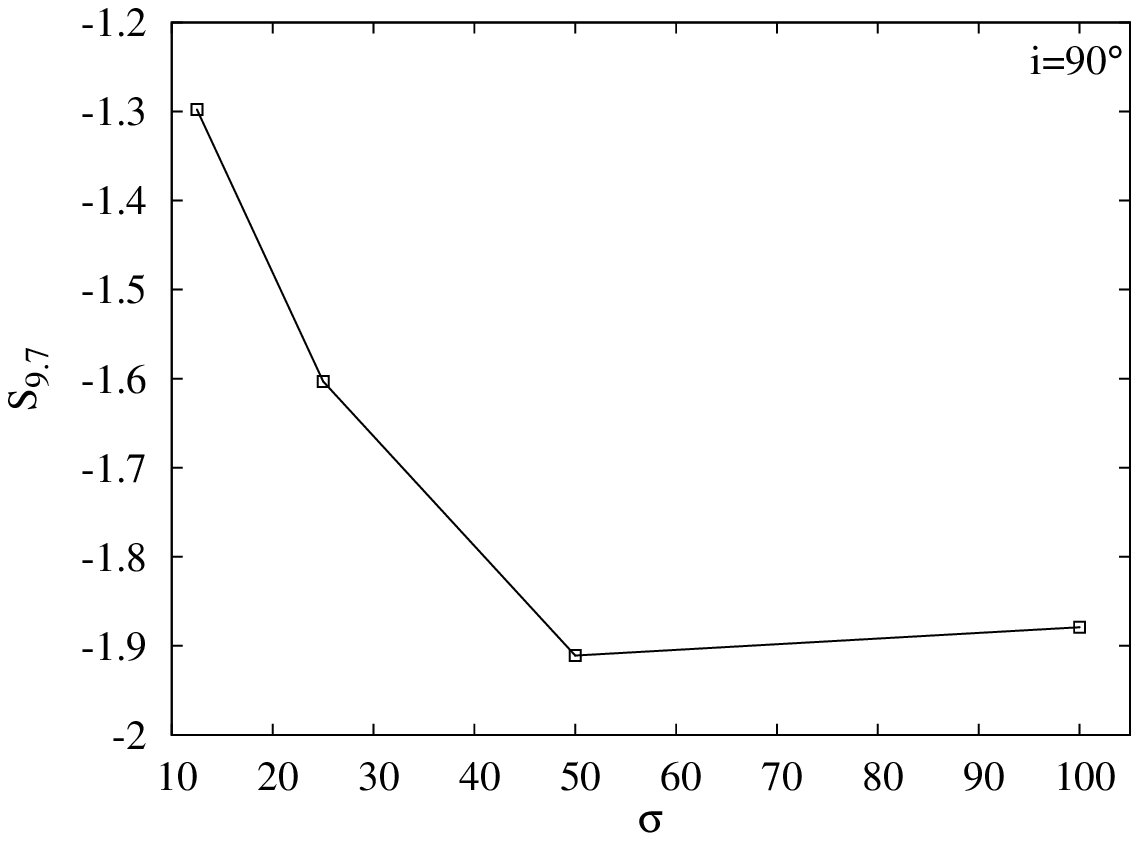}\\
\includegraphics[height=0.3\textwidth]{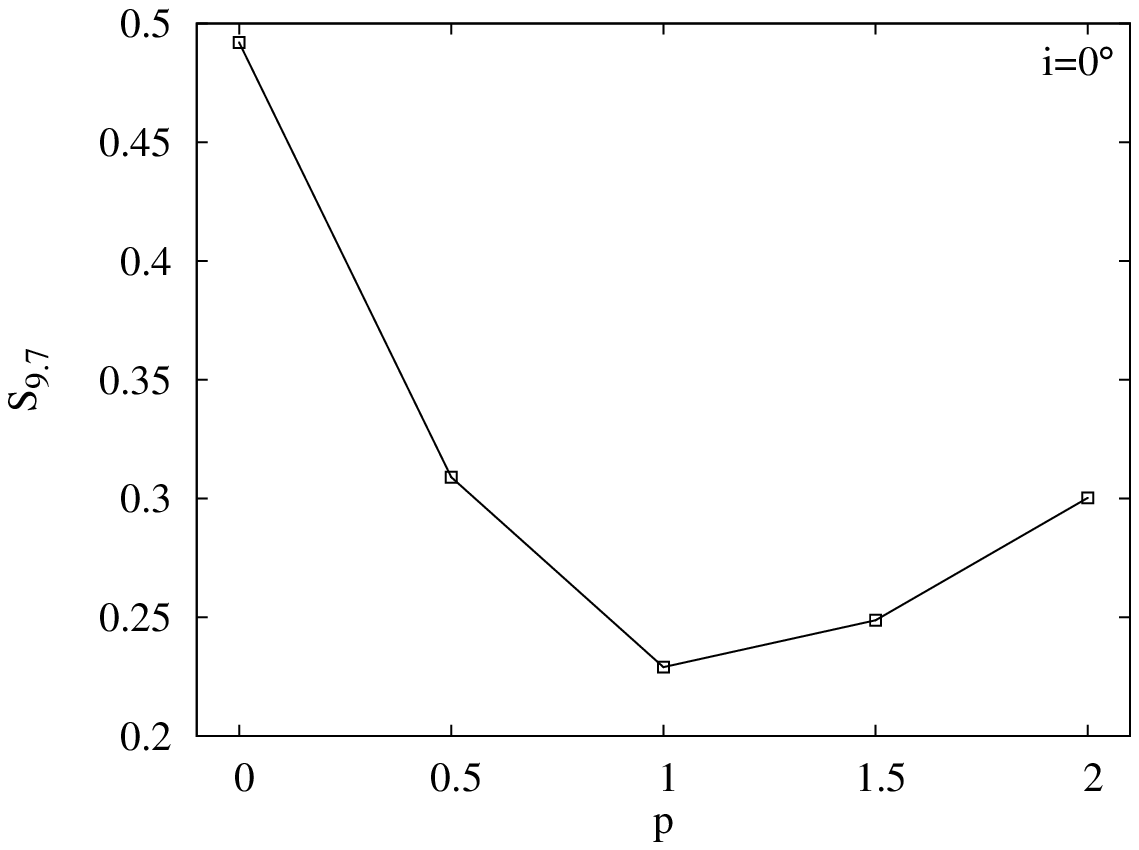}
\includegraphics[height=0.3\textwidth]{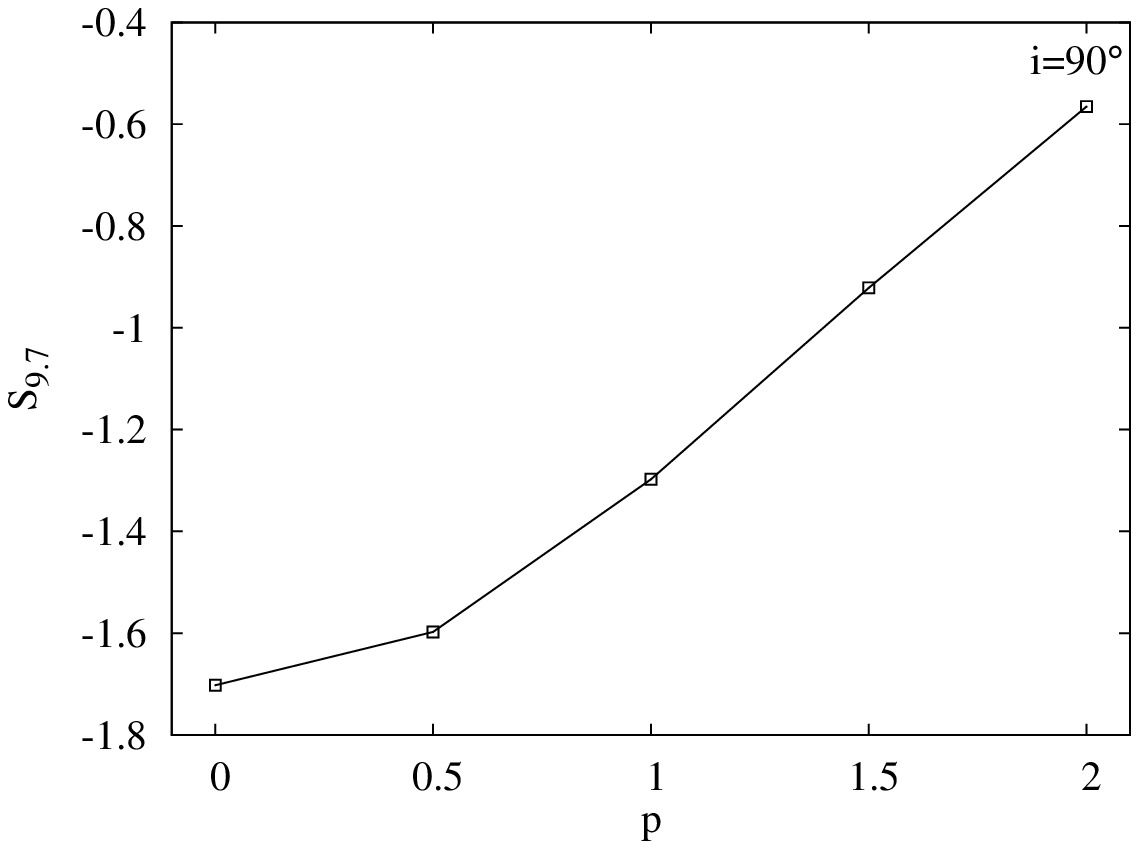}\\
\includegraphics[height=0.3\textwidth]{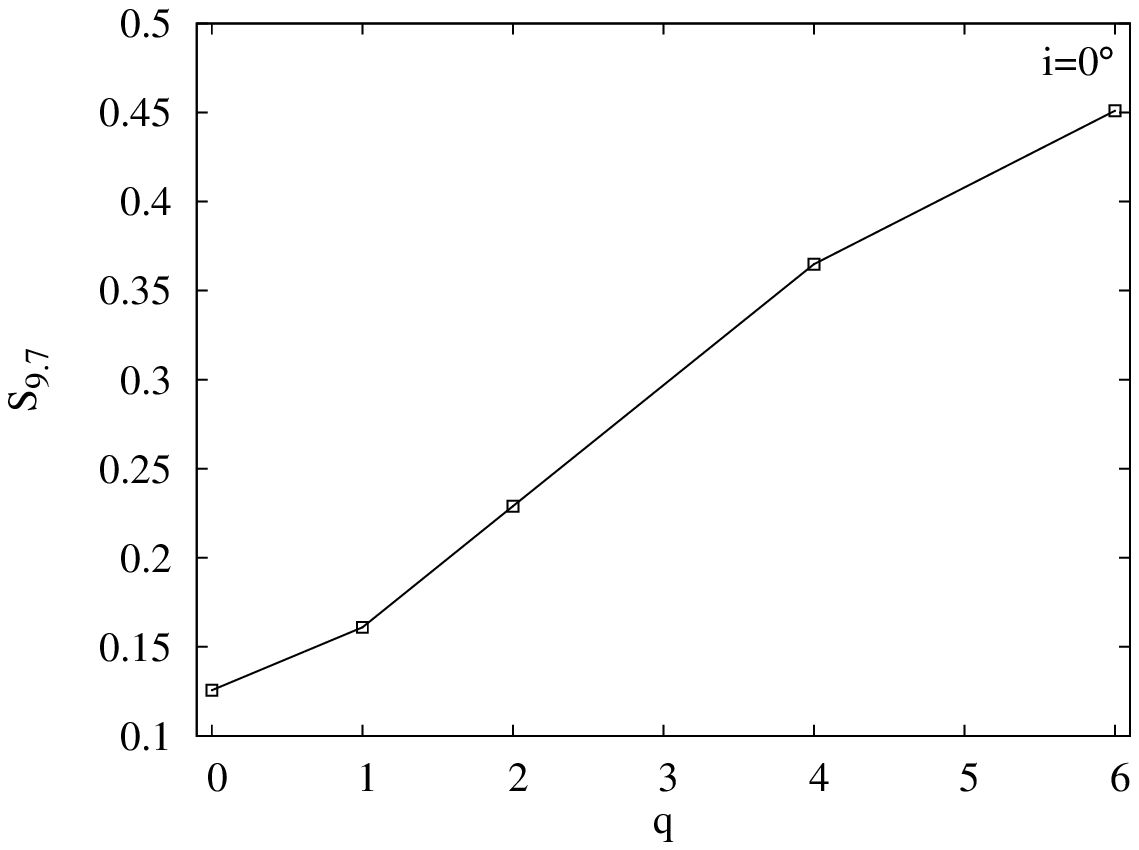}
\includegraphics[height=0.3\textwidth]{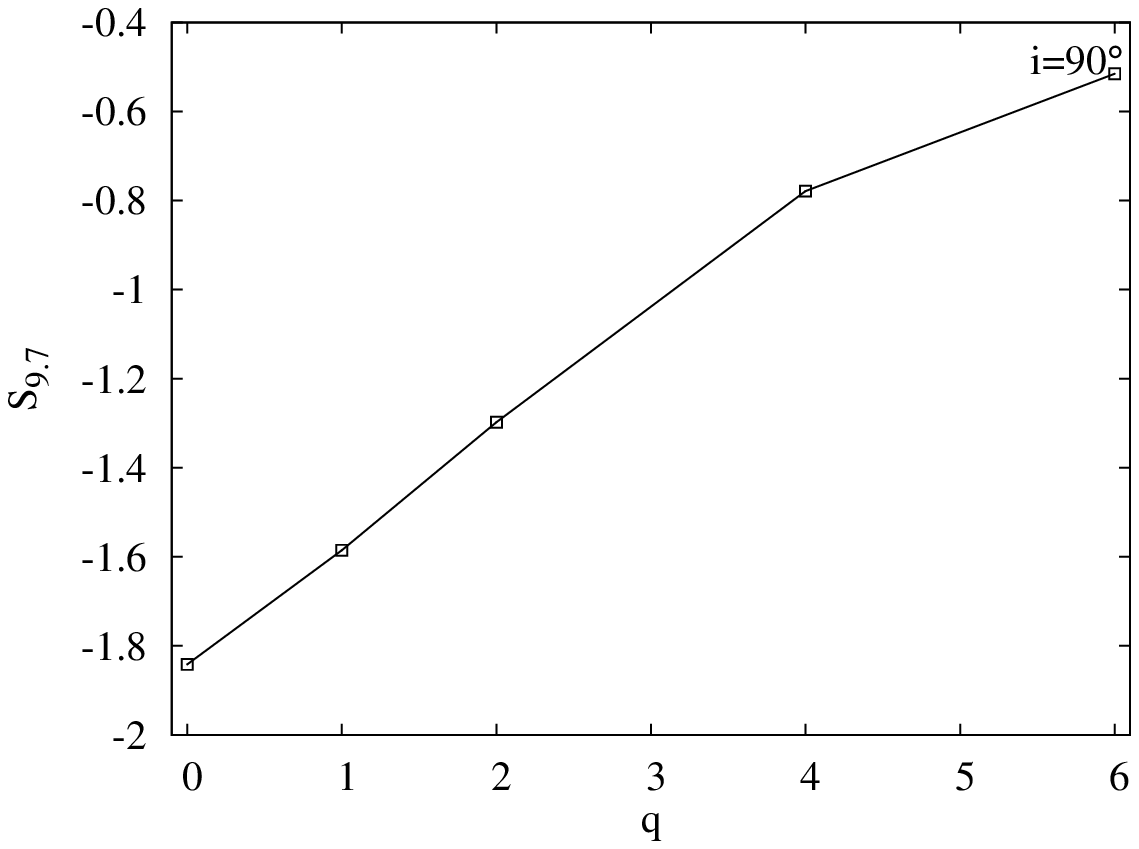}
\caption{Dependence of the strength of the silicate feature ($S$) on
the different parameters. From top to bottom, panels illustrate
dependence on the optical depth ($\tau_{9.7}$), clump size
($\sigma$) and dust distribution parameters ($p$ and $q$).
Panels on the left present values for the face-on view, panels on the
right for the edge-on view.} \label{fig:S}
\end{figure*}

For the majority of the edge-on models with optical depths of
$\tau_{9.7}=5$ and $10$ the silicate feature is in absorption, with
$-2.2 \le S \le -0.2$. Models with low optical depths
($\tau_{9.7}=0.1$ and $1$) do not provide enough dust to absorb the
silicate feature and, in this case, it is present in emission even in
the edge-on view. Fig. \ref{fig:tau} shows SED dependence on the
optical depth. To further illustrate dependence of the strength
of the silicate feature on the different parameters, in 
Fig.~\ref{fig:S} we plot its intensity, $S$, as a function of the
optical depth ($\tau_{9.7}$), of the dust distribution parameters ($p$ and
$q$) and of the clumps size $\sigma$. For these calculations, the following
values of the parameters were chosen: $\tau_{9.7}=10$, $p=1$, $q=0$,
$\sigma=12.5$ and then each of these parameters was varied while all the
others were kept constant.

%---------------------------------
%---------------------------------

%---------------------------------
%---------------------------------
%--------------------------------------------------------------------
\subsection{SED width}
\label{sec:wid} 
%--------------------------------------------------------------------
Following \citet{pierkrolik92} and \citet{granatodanese94}, the width
of the SED, $W$, is defined as the logarithmic wavelength interval in
which the power $\lambda F_\lambda$ emitted in the infrared is more
than one third of the peak value. For a black body this parameter has
a value of $\sim 0.7$, while in the observed spectra its value is
always larger than $1.3$. The vast majority of model SEDs, both in
the face-on and edge-on views, have a width spanning the $1.2 \le W
\le 1.7$ range. SEDs with widths $W>1.55$ are produced by models with
optical depths of $5$ and $10$ because (a)\ these are the models with
the larger amounts of dust and (b)\ the high values of the optical
depth provide a better shielding of the primary source, allowing
colder dust temperatures causing, in turn, the broadening of the SED.
A small fraction of model SEDs have $W<1.15$. These widths are almost
exclusive to models with optical depth of $0.1$ and a density law
parameter $p=1$, which produce the silicate feature in very strong
emission. Since the maximum of the infrared emission often coincides
with the peak of the silicate feature, such models produce lower $W$
values.

Another parameter that affects the SED width is the size of the
torus. Increasing the radius, while keeping the optical depth
constant, means that the amount of dust in the outer (and colder)
regions increases. As these regions emit in the far-infrared, an
increase in the radius makes the SED wider. For the same reason, $W$
will increase with the total amount of dust, that is, with the
optical depth (see Fig. \ref{fig:tau}).

The edge-on orientations produce wider SEDs than the face-on ones,
with almost 50\% of them having $W>1.6$. This is because in the
edge-on view the silicate feature is usually in absorption. As a
result, the peak of the infrared emission decreases, leading to a
wider SEDs. Furthermore, in the edge-on view the received
radiation is mainly coming from the outer regions that contribute to
the far-infrared emission.

%--------------------------------------------------------------------
\subsection{Isotropy of the infrared emission}
\label{sec:iso}
%--------------------------------------------------------------------

Following \citet{dullemond05} we define the isotropy parameter, $I$,
as the ratio of the total integrated infrared flux in an edge-on view
over the total integrated infrared flux in a face-on view. Larger
value of $I$ implies there is more isotropy.

Anisotropy in the infrared emission is expected in all systems with
torus-like geometry. This is because in the face-on view we have a
direct view of the primary source and the inner, hotter region of
the torus, while in the edge-on view they are obscured. The
values of $I$ strongly depend on the optical depth. Models with a low
optical depth are almost perfectly isotropic: models with
$\tau_{9.7}=0.1$ produce $I>0.95$ and in models with $\tau_{9.7}=1$
$I$ takes values around $\sim 0.75$. Models with a higher optical
depth have anisotropic emission, with most $I$ values being around
$\sim 0.50$ and $\sim 0.40$ for optical depths of $5$ and $10$,
respectively. The lowest $I$ value in our models is $\sim 0.37$.

%--------------------------------------------------------------------
\subsection{The peak of the infrared emission}
%--------------------------------------------------------------------
Another important feature characterizing the infrared SED of AGNs is the 
wavelength at which it peaks. We measure this quantity in our model
SEDs expressed in $\lambda F_\lambda$. The majority of the models in
our grid peak around $\sim 9.4\, \mu$m, more or less corresponding to
centre of the silicate band. A small fraction of models has its
maximum in the $\sim 20$ to $\sim 29\,\mu$m range: all of these
models have a high optical depth (either 5 or 10). In the face-on
view, almost all models peak at $\lambda=9.4\,\mu$m. In the edge-on
view, models exhibiting the silicate feature in emission (i.e. models
with low optical depths of $\tau_{9.7}=0.1,1$) also peak at $9.4\,
\mu$m, due to the prominence of the $10\,\mu$m feature in emission
in low-optical depth systems and lower dust content. Edge-on models
with higher optical depths peak beyond $\sim 20\,\mu$m. 

%--------------------------------------------------------------------
\subsection{Comparison of two-phase and smooth models}
\label{sec:smooth}
%--------------------------------------------------------------------

%---------------------------------
\begin{figure*}
\centering
\includegraphics[height=0.34\textwidth]{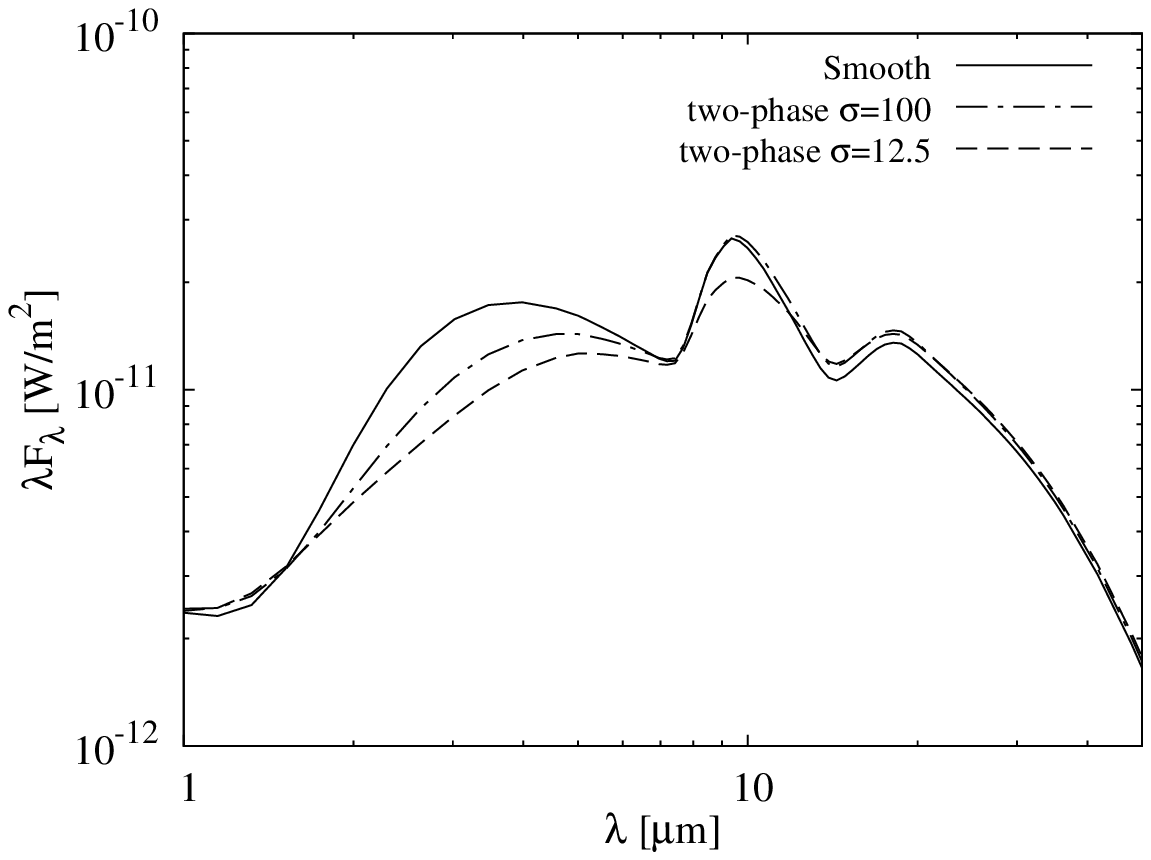}
\includegraphics[height=0.34\textwidth]{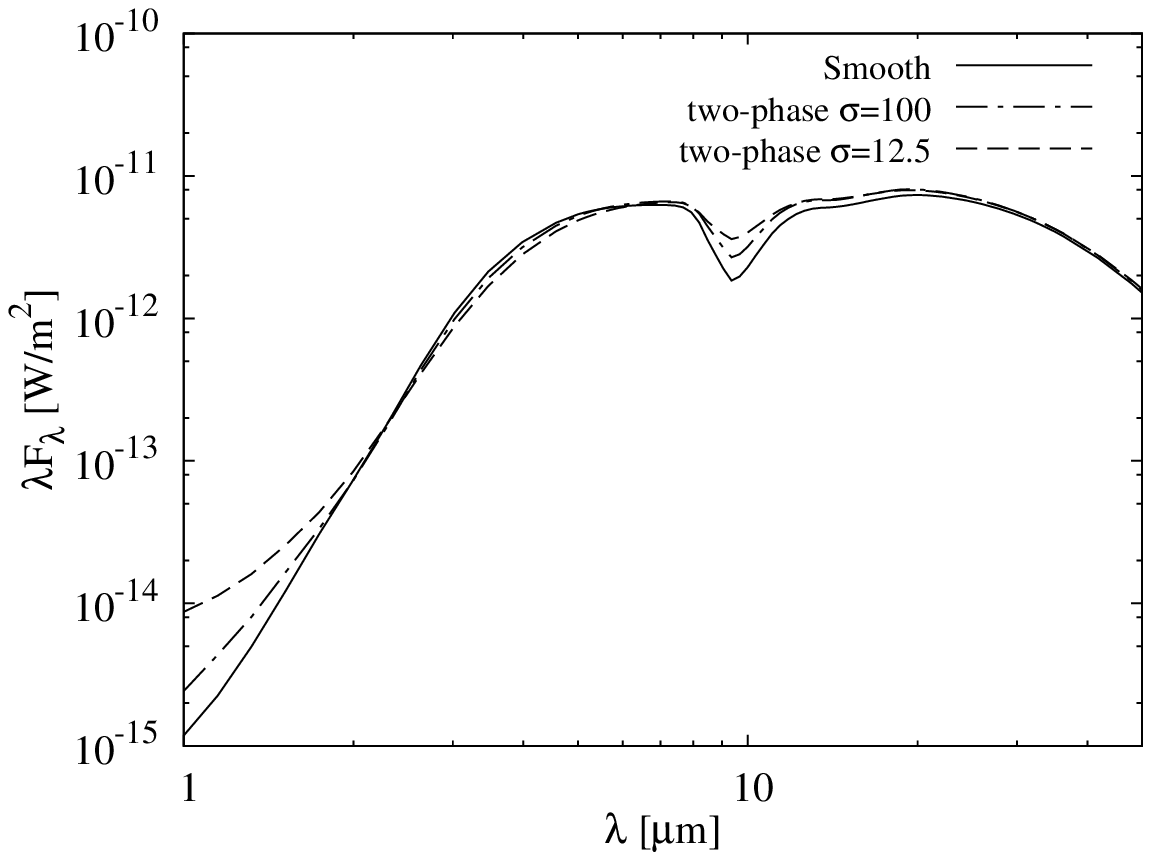}
\caption{Comparison of smooth and clumpy model SEDs. Full solid line
represents smooth model, dotted line two-phase model with
$\sigma=100$, dashed two-phase model with $\sigma=12.5$ and
dashed-dotted clumps-only model with $\sigma=12.5$. All other
parameters are the same as in Fig. \ref{fig:inc}. Left panel: face-on
view; right panel: edge-on view.}
\label{fig:clump_smooth}
\end{figure*}
%---------------------------------

%---------------------------------
\begin{figure*}
\centering
\includegraphics[height=0.6\textwidth]{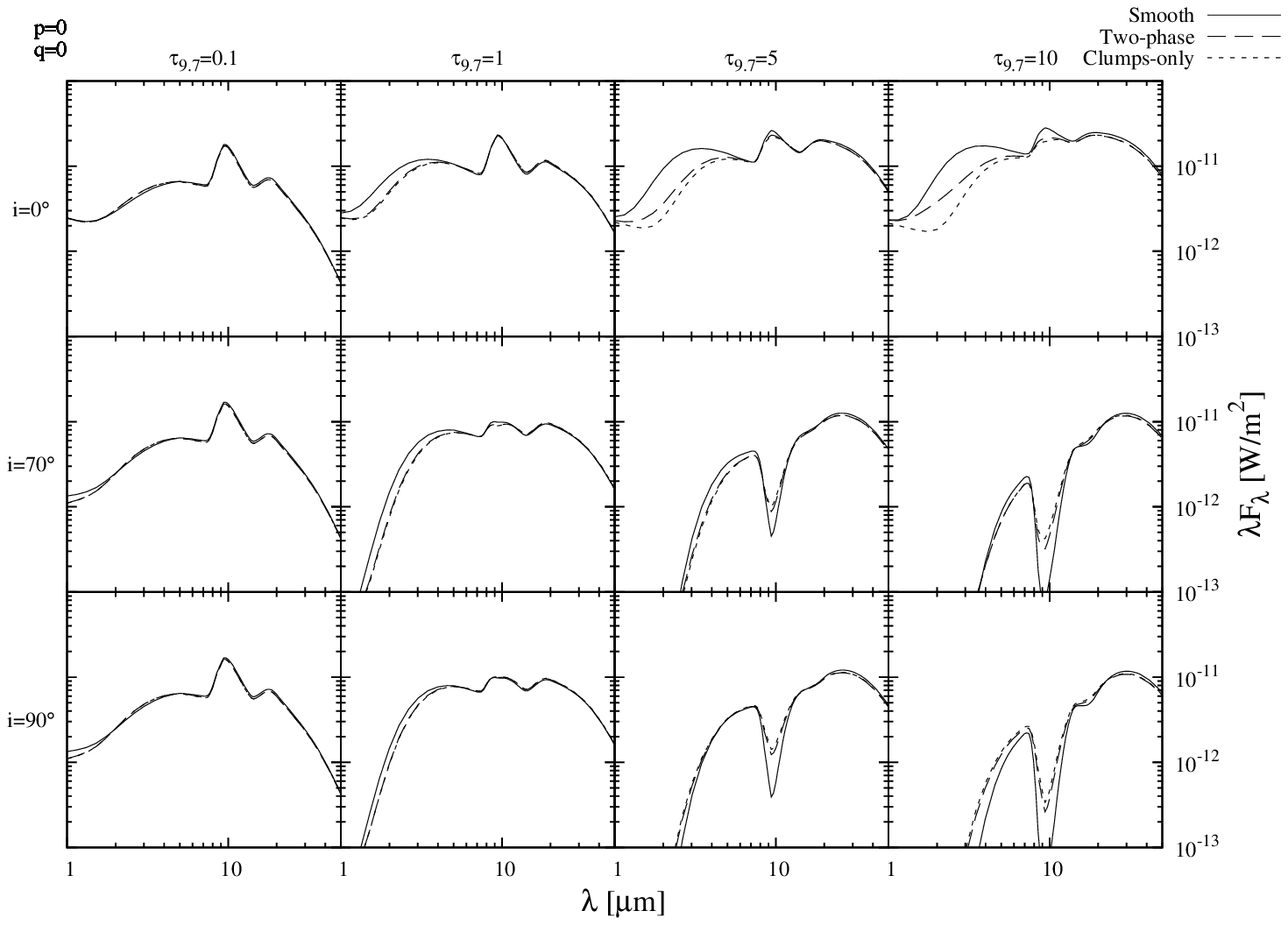}
\includegraphics[height=0.6\textwidth]{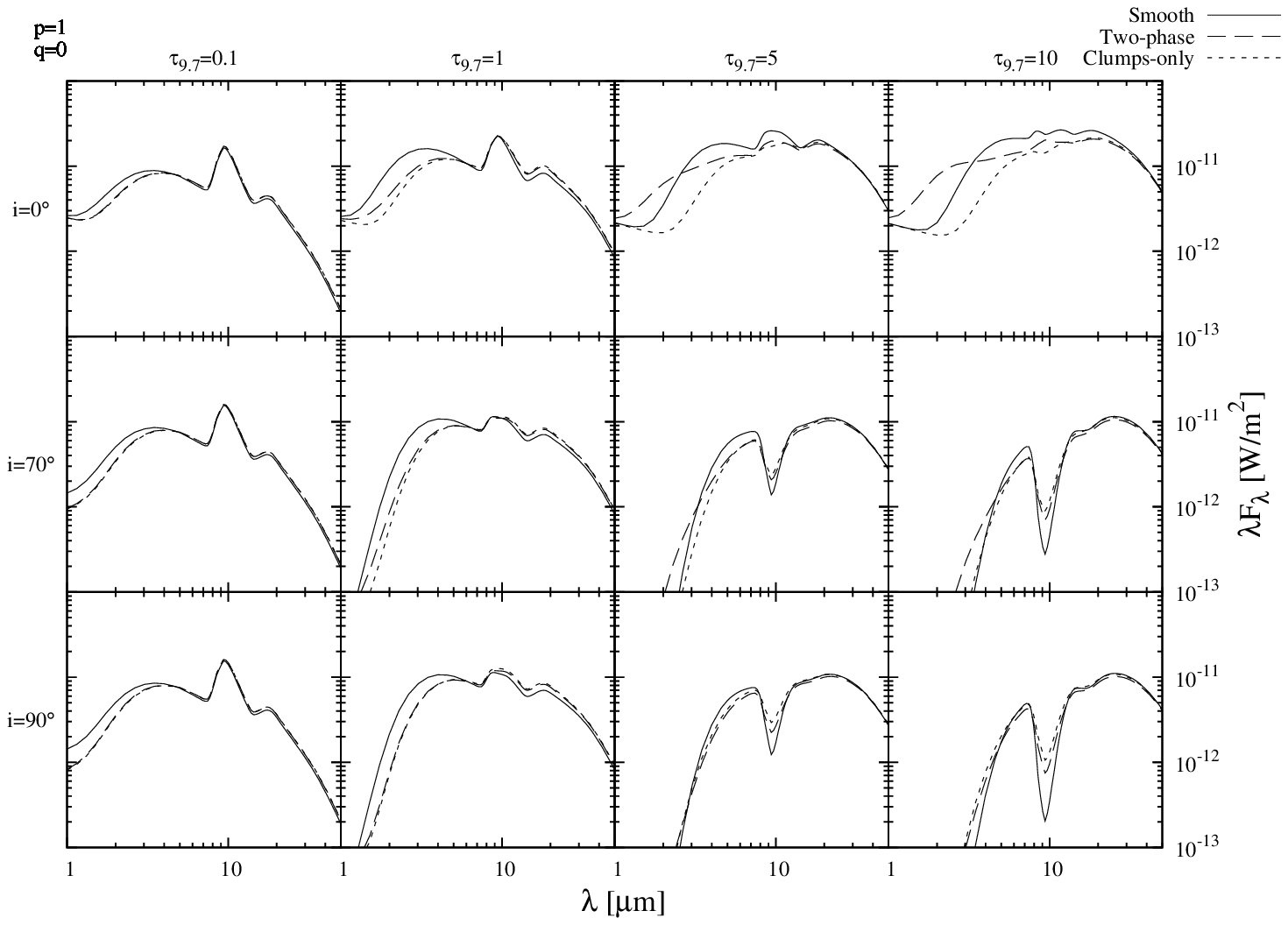}
\caption{SEDs of our standard model grid, in the $1-50$ $\mu$m
wavelength range. Solid line: smooth models; dashed line: two-phase
models; dotted line: clumps-only models. The columns correspond to
optical depths of $\tau_{9.7}=0.1,1,5,10.0$, from left to right. The
rows correspond to inclinations of $i=0,70,90^\circ$, from top to
bottom. The dust distribution parameters ($p$ and $q$) are given
in the top left corner of each panel.}
\label{fig:sedgrid}
\end{figure*}
%---------------------------------
%
\addtocounter{figure}{-1}
\begin{figure*}
\centering
\includegraphics[height=0.6\textwidth]{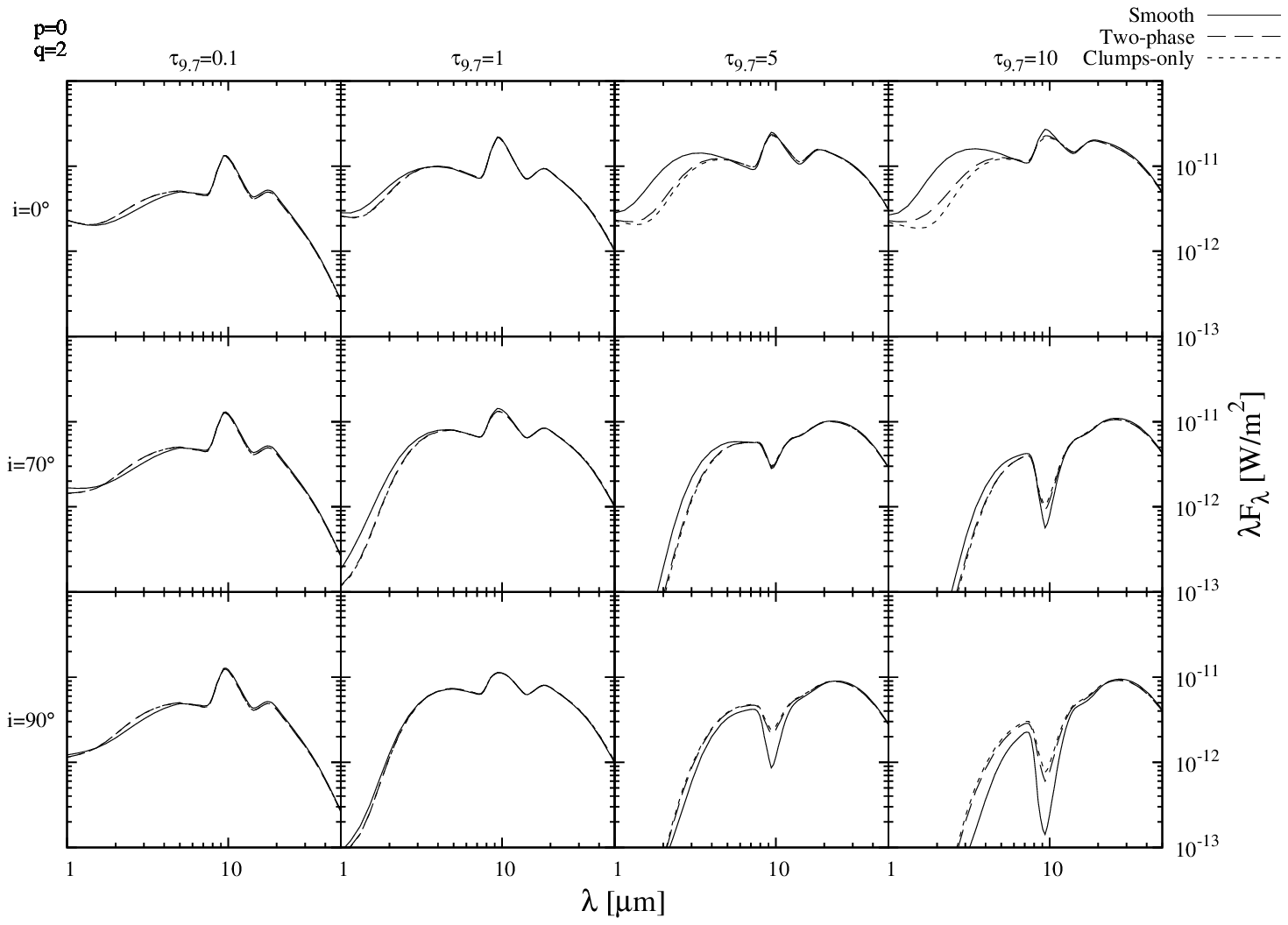}
\includegraphics[height=0.6\textwidth]{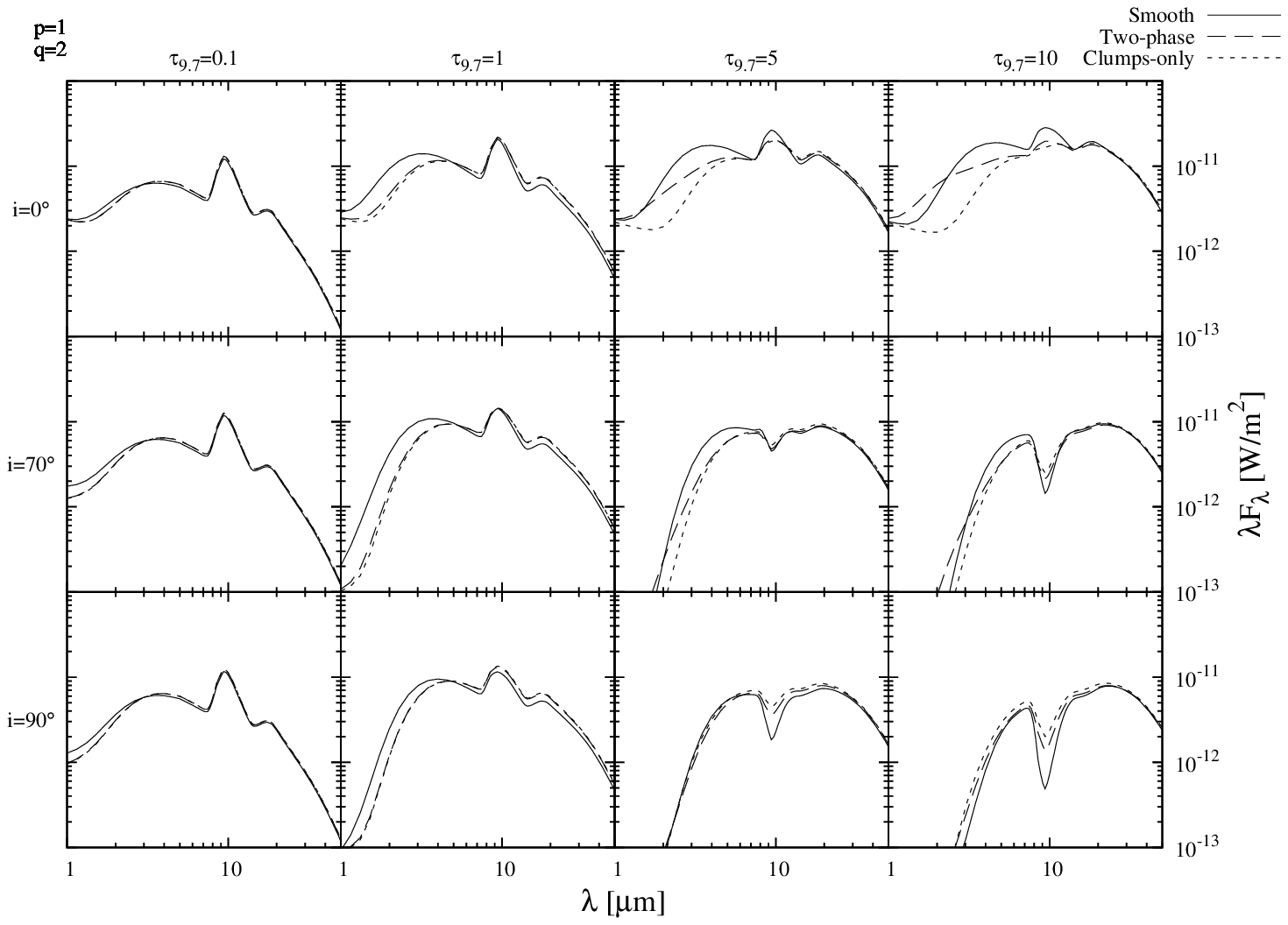}
%\contcaption
\caption{- \textit{continued}}
\end{figure*}
\addtocounter{figure}{-1}
\begin{figure*}
\centering
\includegraphics[height=0.6\textwidth]{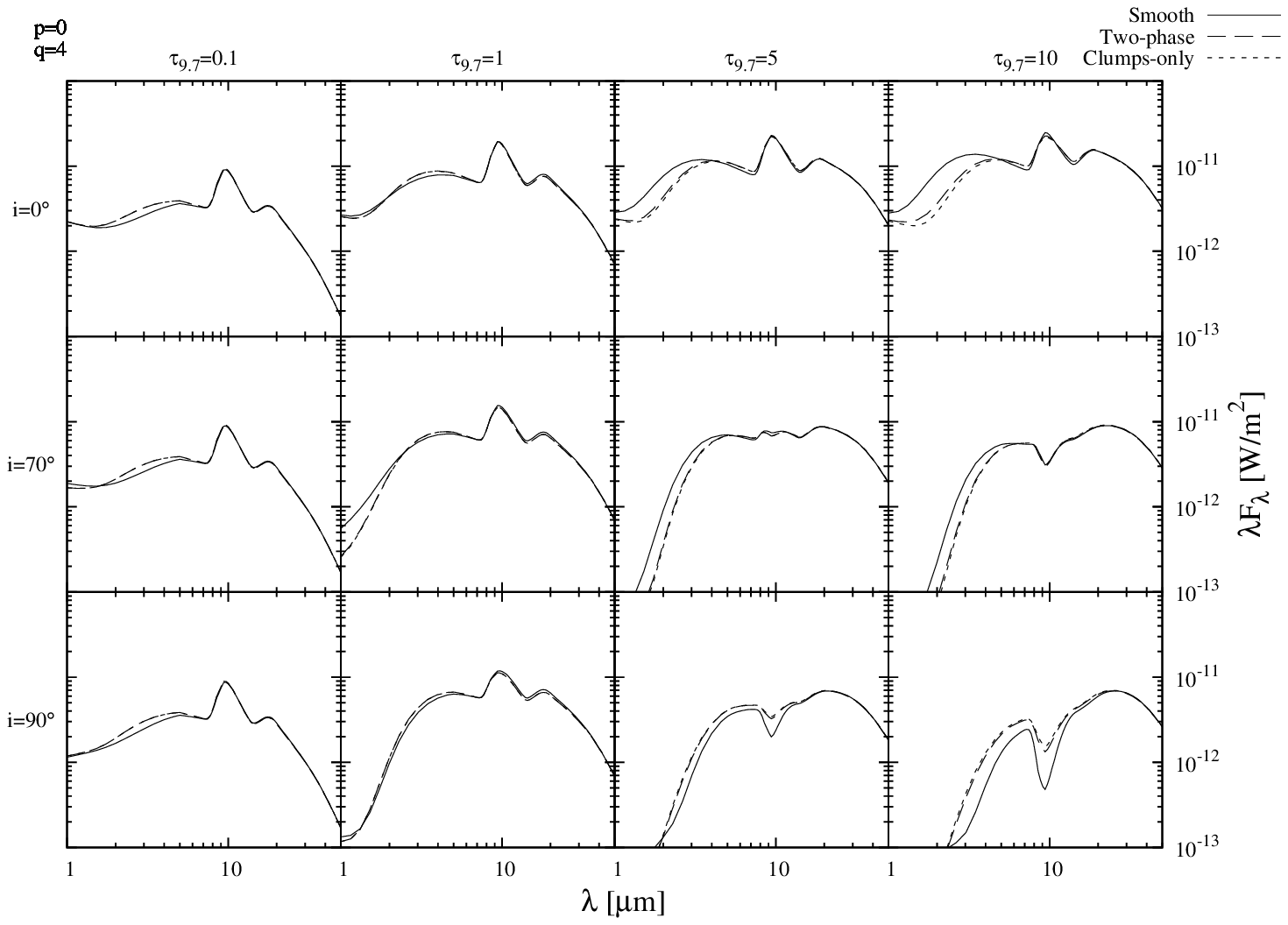}
\includegraphics[height=0.6\textwidth]{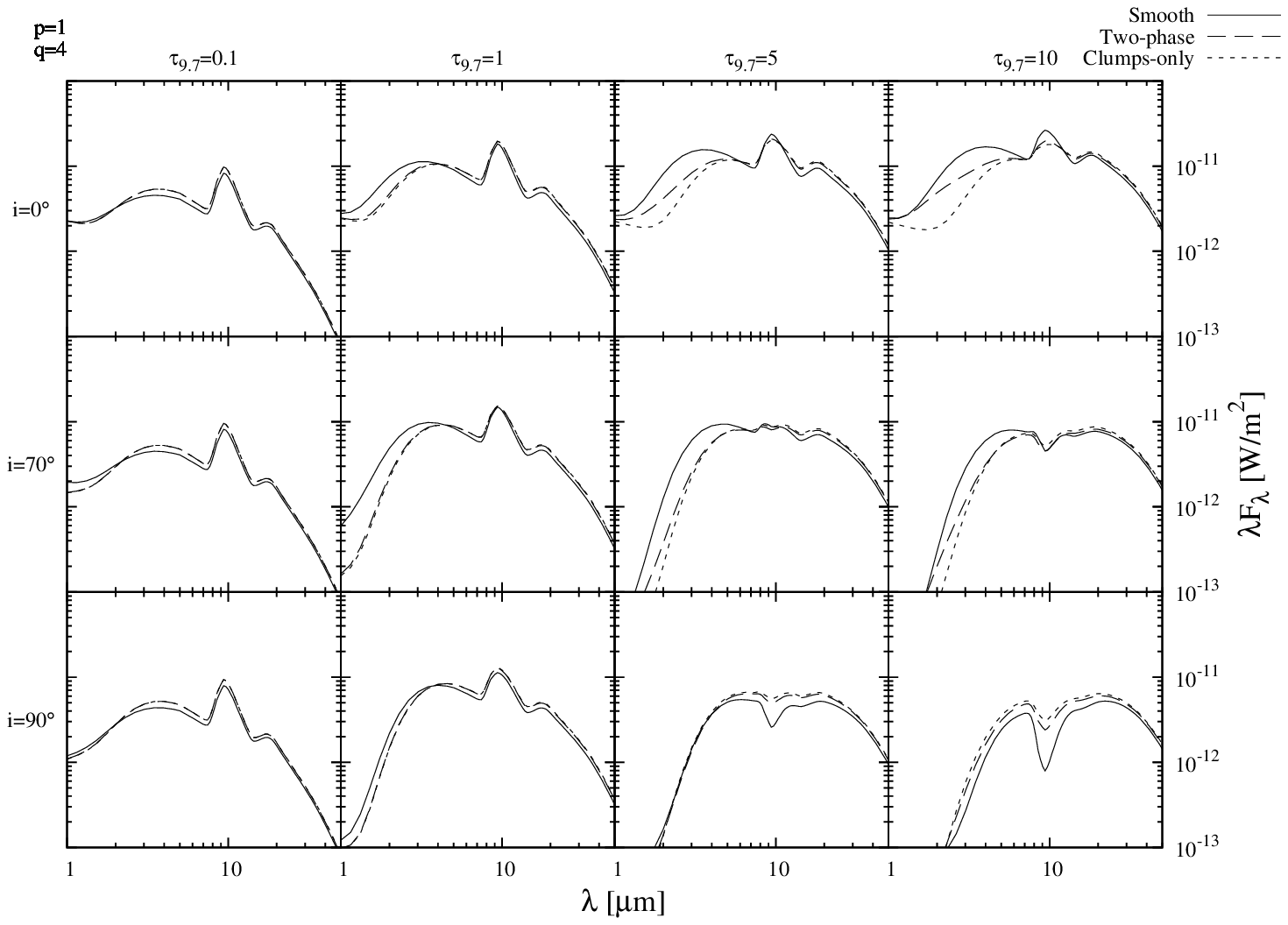}
%\contcaption{}
\caption{- \textit{continued}}
\end{figure*}
\addtocounter{figure}{-1}
\begin{figure*}
\centering
\includegraphics[height=0.6\textwidth]{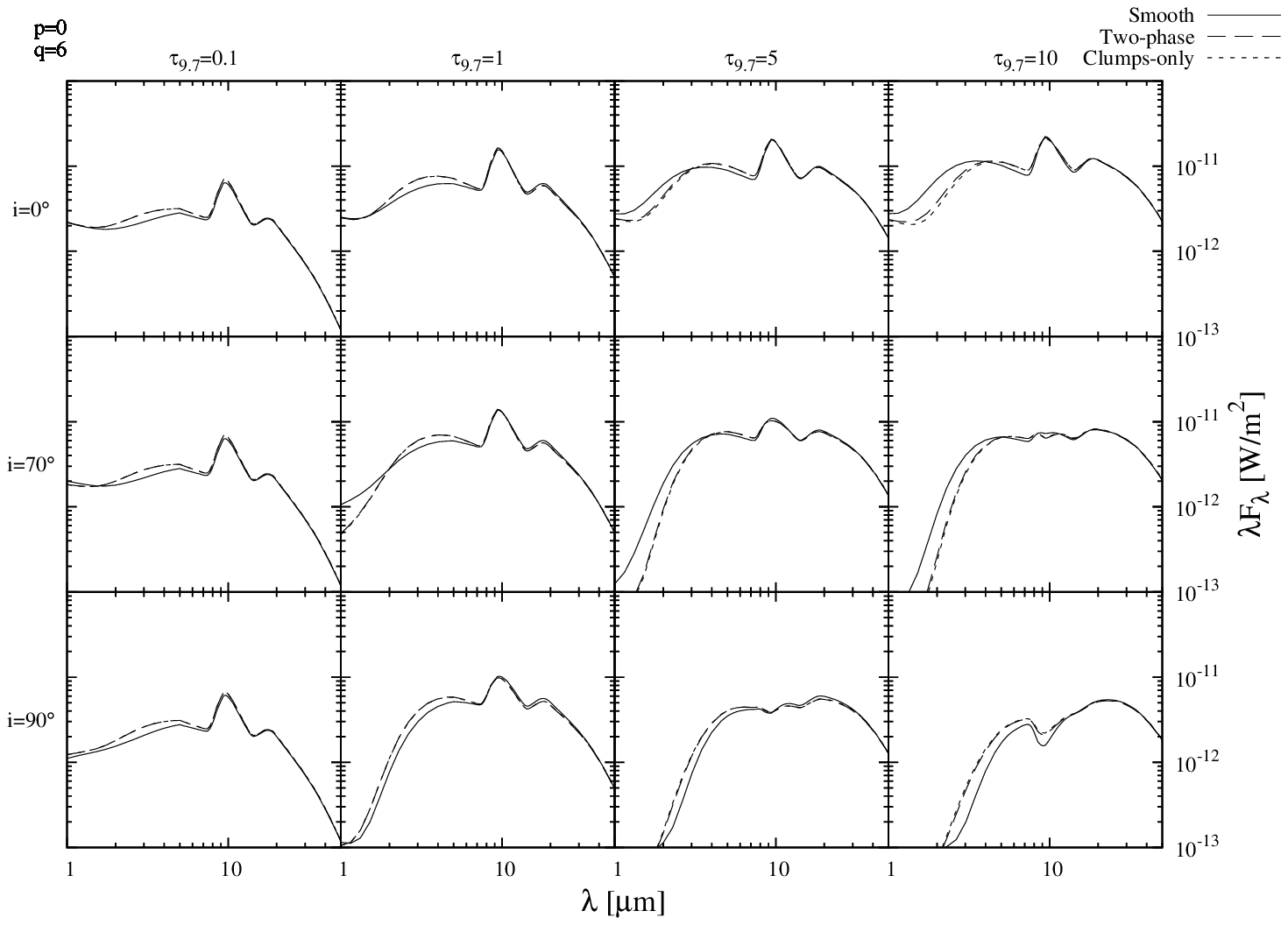}
\includegraphics[height=0.6\textwidth]{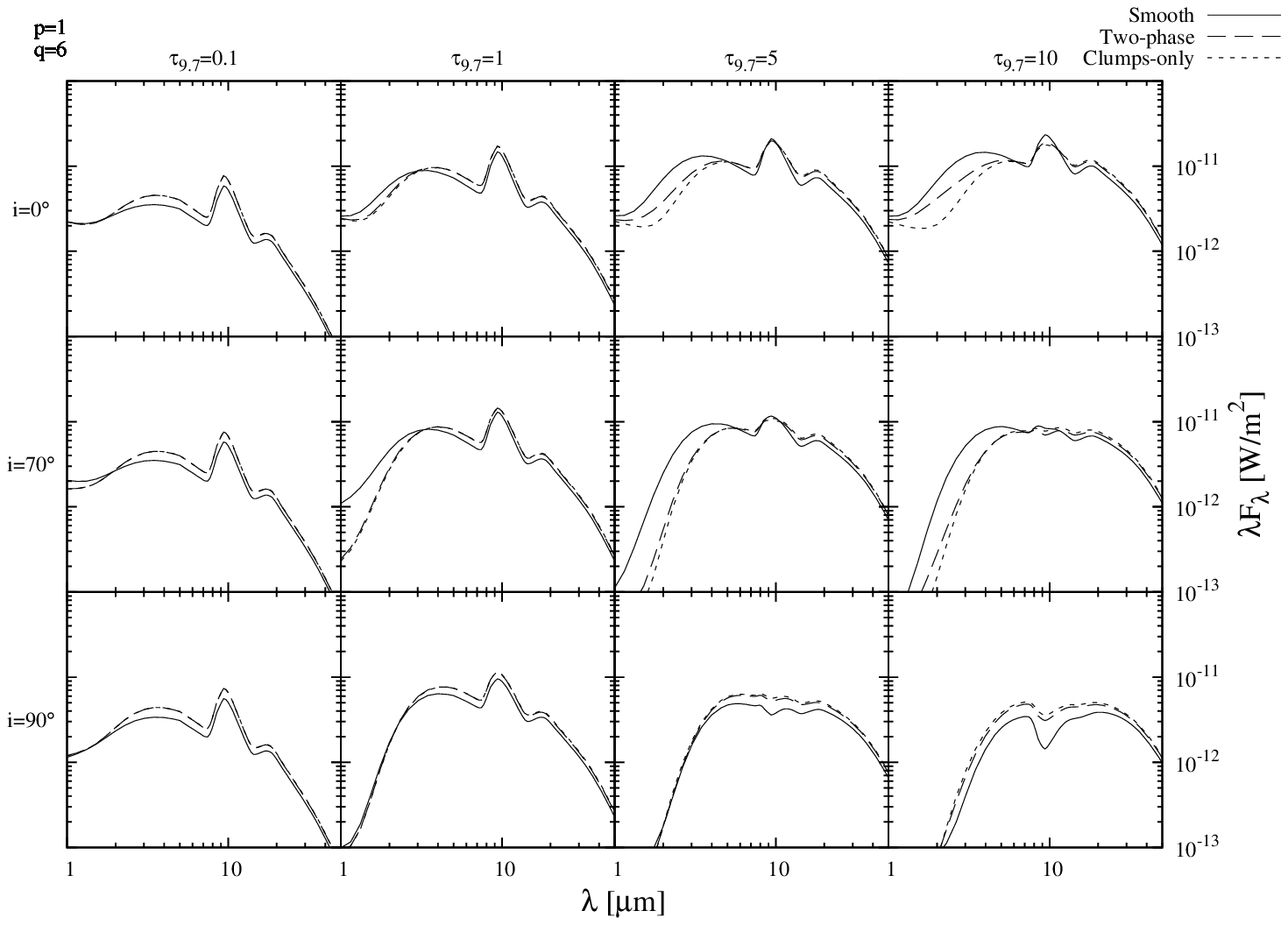}
%\contcaption{}
\caption{- \textit{continued}}
\end{figure*}
%
%--------------------------------------------------------------------

In this section we investigate the differences between the models
with homogeneous dust distribution (smooth models) and models with
dust as a two-phase medium. In order for this comparison to be as
consistent as possible, for each two-phase model we have generated
its corresponding smooth configuration, using the same global
physical parameters. Furthermore, we have generated two different
sets of two-phase models using a relative clump size (see Eq.
\ref{eqn:size}) value of $\sigma=100$ and $\sigma=12.5$,
respectively: in the latter case, the clumps are eight times bigger
than in the former.

We found that two-phase models with $\sigma=100$ (small clumps) tend
to have a less pronounced emission in the $1-6\,\mu$m range, when
compared to the smooth ones. If we compare the intensity of the
$10$ $\mu$m silicate feature, we find virtually no difference in
type 1 view, while slightly lower absorption are measured in
two-phase SEDs for type 2 lines of sight. As expected, the dust
distributed in a large number of small clumps, embedded in a smooth,
homogeneous medium, will closely resemble the characteristics of a
smooth SED.

Two-phase models with bigger clumps ($\sigma=12.5$) are showing more
difference compared to both smooth and $\sigma=100$ models. In the
face-on view, they tend to have even less pronounced emission and
also a different, flatter, slope in the $1-6$ $\mu$m range. Depending
on the parameters of the dust distribution and on the optical depth,
the silicate feature is in general less pronounced. This
behavior can be attributed to the shadowing effect caused by the
clumps in the innermost region, where the dust is hotter and the
feature is produced. In the edge-on view, the silicate band
absorption is less deep compare to both smooth and
$\sigma=100$-models because we are able to penetrate -- at least
partly -- between the clumps deeper into the torus. Fig.
\ref{fig:clump_smooth} presents a comparison of SEDs of typical
models with the smooth and the two two-phase dust distributions for
the two clumps sizes.

For face-on view, although both smooth and two-phase
models are able to produce almost the same range of values of the
silicate feature strength, two-phase models tend in general to
produce attenuated emission compared to those produced by the
corresponding smooth models. The majority of both smooth and
two-phase model SEDs have their infrared emission maximum around
$\sim 9.4$ $\mu$m. However, while no smooth model peaks beyond $12$
$\mu$m, there are several two-phase models that peak around $\sim 20$
$\mu$m. This is because the two-phase models tend to produce an
attenuated silicate emission feature, and when it is very weak or
absent, the peak of the emission is shifted toward the longer
wavelengths. In the edge-on view, the two-phase models produce a
weaker silicate absorption feature, with the lowest strength around
$-2.4$. The smooth models produce a deeper silicate feature, with the
strength value reaching a minimum of $-4.4$. 

Two more characteristics which are of interest when comparing smooth
and two-phase models, are the isotropy of the infrared emission and
the SED width (see sec. \ref{sec:iso} and \ref{sec:wid} for
definitions). Both two-phase and smooth models produce a similar
range of values of the isotropy parameter $I$. However, compared
individually, two-phase models are more isotropic than the smooth
ones. Regarding the SED width, $W$, we found that clumpiness does not
have a profound effect on this parameter.

In Fig. \ref{fig:sedgrid} we present plots of SEDs covering our
standard parameter grid, for three characteristic inclinations
($0^\circ$, $70^\circ$, $90^\circ$). This figure illustrates how SEDs
of smooth, two-phase and clumps-only models compare to each other and
evolve with the different parameters, i.e. inclination, optical depth
and the two parameters determining the dust distribution. In models
with low optical depth, the silicate feature appears in a strong
emission and the difference between smooth and clumpy models is
marginal. With increasing optical depth the difference is increasing
as well. Also, the difference between smooth and clumpy models is
greater in the cases of constant dust density with polar angle ($q=0$
in Eq. \ref{eqn:dens}), and non-constant dust density in the radial
direction ($p=1$).

%--------------------------------------------------------------------
\subsection{Comparison of two-phase and clumps-only models}
\label{sec:clonly}
%--------------------------------------------------------------------
As it can be seen from Fig. \ref{fig:sedgrid}, the major difference
between SEDs of two-phase and clumps-only models arises in the
near-infrared range and mainly for face-on view. At these
wavelengths, most of the two-phase models with type 1 inclination
have a flatter SED when compared to the corresponding clumps-only
models. This difference is caused by the presence of the smooth
component in which the clumps are embedded. Dust in this component,
exposed to the radiation field of the central source, can reach high
temperatures and will give rise to higher luminosity in the $2-6\,
\mu$m range.

Regarding the $10$ $\mu$m silicate feature, we do not find any
significant difference between the two dust configurations: depending
on the parameters, in clumps-only models it could be slightly
attenuated compared to the one in the two-phase models, but the
difference is in the most cases marginal. A similar behaviour can be
observed in SEDs of edge-on views, in which the smooth low density
component is responsible for additional absorption, so the silicate
feature is slightly deeper in the two-phase models. The
dissimilarities between the SEDs in these two dust configurations
increase as the optical depth increases: from models with the lowest
value ($\tau_{9.7}=0.1$), where the SEDs are identical, to models
with the highest value ($\tau_{9.7}=010$) which display the most
evident differences. The difference is the most pronounced in the
cases of constant dust density with polar angle ($q=0$), and
non-constant dust density in the radial direction ($p=1$). 

It is very interesting to note that such a behaviour of the
near- and mid-infrared SED of the two-phase dust distribution, would
overcome an issue that seems to be common to the most clumpy models
currently available in the literature. Exploiting the model of
\citet{nenkova08b}, \citet{mor09} fit a sample of mid-infrared
spectra
of 26 luminous quasar, finding the need of an extra hot-dust
component, which they add to the clumpy torus SED, in order to
properly reproduce the shorter wavelengths part of the
\textit{Spitzer}
spectrum. The addition of this hot dust, whose emission is
represented by a black-body with a temperature of about the
sublimation limit of graphite, is required by the lack of emission
from the adopted clumpy model at these wavelengths. More recently
\citet{deo11}, adopted the same clumpy model to reproduce a
combination of observed broad-band photometry and the mid-infrared
spectrum of 26 high redshift type-1 quasars. Similarly to
\citet{mor09}, the adopted clumpy models are not able to
simultaneously reproduce the intensity of the silicate feature and
the near-infrared continuum emission: models that would properly fit
the continuum were overestimating the silicate feature emission. An
analogous problem was also spotted by \citet{vignali11}, when using
the same clumpy models to fit the observed photometry and IRS
spectrum of a $z\sim 0.44$ type-2 quasar. Adopting the clumpy models
developed by \citet{honig06}, to fit both photometry and mid-infrared
spectroscopy data, \citet{polletta08} reach similar conclusions.

As we have shown, a torus model with the dust distributed in a
two-phase medium, has a more pronounced (`hotter') emission in the
$2-6$ $\mu$m range while displaying, at the same time, a silicate
feature whose intensity is almost identical to that of the
corresponding clumps-only model.

\subsection{Other results in the literature}
\label{sec:comparison}
Making a detailed comparison of our modeling approach with
models 
previously developed in the literature and their results, is quite a
tricky
task, and is beyond the scope of our work. Furthermore, what we
describe 
in this paper is a model which would ideally put itself in between
smooth
and clumpy models approach, and it is hence not directly comparable
to
any of formerly published work. In this section we give a very brief
description,
which is by no means meant to be exhaustive, of some of the
aforementioned
works, limiting ourselves to models that consider a clumpy dust
distribution.

The exploitation of radiative transfer codes to model AGN IR
emission, taking 
into account the clumpy nature of dust surrounding the central
source, 
includes at present only few works:
\cite{nenkova02,nenkova08a,nenkova08b}, 
\cite{dullemond05}, \cite{honig06,honig10}, \cite{schartmann08} and
 \cite{kawmor11}. Each of these works exploits different techniques 
 and approximations.

In their series of works, Nenkova et al. used the radiative transfer
code 
\texttt{DUSTY} \citep{ivezic97} to solve the radiative transfer
equation 
for the single clouds, that where modeled as 1-D dust slab. The
final 
torus SED was obtained by adding the emission from different slabs
at 
different viewing (phase) angles, after statistically weighting them.
They 
find that 5 to 15 clumps in a  equatorial line of sight, each with an
optical 
depth in the range $\tau_V \sim 30-100$, are successful in
reproducing 
the observed characteristics of AGNs. Models in which the clouds are
more
concentrated at shorter distances from the central source, i.e. with
a radial
distribution following a power low, are favoured.

Another approach was followed by \cite{dullemond05}, who exploit a
2-D 
Monte-Carlo code, in which the clumps where modeled as concentrical 
rings. Once the temperature of the dust is known throughout all the
cells, 
the torus SED is calculated by means of ray tracing techniques.
Starting 
from models with dust continuously distributed, they calculate the
respective 
clumpy models, finding that it is not possible to use observed
infrared data 
to distinguish between the effects due to the two different
distributions.

The model developed by \cite{honig06} and its further development 
\citep{honig10}, also adopts a Monte-Carlo technique to solve the
radiative 
transfer problem, calculating the SEDs for various phase angles, for 
each cloud, setting up in this way a database of clumps emission.
They 
consider that both the clouds optical depth and their size (radius)
are related
 to their distance from the central source. The clouds are then
randomly 
 displaced, according to a spatial distribution function, and the
torus SED 
 is calculated by summing the emission of directly and non-directly
illuminated 
 clumps. This approach allows them to also study the dependence of
the dust 
 SED on the arrangement of the clouds: relevant differences are
found 
 especially for intermediate angle lines of sight. 

Monte-Carlo, coupled to ray tracing techniques, are used by
\cite{schartmann08},
 who are not using any prescription for the dust distribution which
is instead 
 computed from the equilibrium between the gravitational potential
and pressure 
 forces. They explore the effect of the clouds filling factor, of
changing the dust mass, 
 of the clump size and their positions. Again, their analysis of the
SED for different 
 arrangements of the clumps, shows non-negligible differences which
tend to be 
 the highest for edge-on views. The case of a non--isotropically
emitting central 
 source, whose emission is varying according to a $| \cos(\theta)|$
law, was also
 studied by \cite{schartmann05}, but their results are not directly
comparable to ours
 since in their case the dust was continuously distributed.
  
%--------------------------------------------------------------------
\section{Conclusions}
\label{sec:conc}
%--------------------------------------------------------------------

In this paper we have investigated the infrared emission of AGN dusty
tori. Following theoretical predictions coming from hydrodynamical
simulations, we modeled the dusty torus as a 3D two-phase medium with
high-density clumps and low-density medium filling the space between
the clumps. We employed a 3D radiative transfer code based on the
Monte Carlo technique to calculate SEDs and images of torus at
different wavelengths. We calculated a grid of models for different
parameters and analyzed the properties of the resulting SEDs. For
each two-phase model we have calculated two corresponding models with
the same global physical parameters: a clumps-only model and a model
with a smooth dust distribution. For both two-phase and clumps-only
models, another set is generated keeping all the parameters constant
but varying the random distribution of the clumps. From the analysis
of the SED properties and comparison of the corresponding models, we
conclude the following:

\begin{itemize}

\item The SED at near- and mid-infrared wavelengths is determined by
the conditions of dust the innermost region of the torus: different
random distributions of the clumps may result in the very different
SEDs in otherwise identical models.

\item The shape of the silicate feature is not only a function of
inclination. Optical depth, dust distribution parameters, clump size
and actual arrangement of the clumps, all have an impact on the
appearance of the silicate feature. Low optical depth tori produce
silicate feature in a strong emission. Models with high-density
clumps occupying the innermost region will have the emission feature
attenuated due to the shadowing effects.

\item The clump size has a major impact on the SED properties. SEDs
of the clumpy models with small clumps ($\sigma = 100$ or clump size
of $0.15$ pc) are very similar to the ones obtained by a homogeneous
distribution of the dust. The silicate feature in absorption
in these models is shallower and they tend to have less near-infrared
emission than the corresponding smooth models. However, the silicate
feature in emission is not suppressed. Clumpy models with bigger
clumps ($\sigma = 12.5$ or clump size of $1.2$ pc) are showing more
differences compared to both small clump and smooth models. The
silicate feature in absorption in these models is even less deep and
they have less near-infrared emission than the small clump and smooth
models. The silicate feature in emission is in general less
pronounced. We stress that suppression strongly depends on the dust
distribution parameters. The effect is the most notable in the case
of a non-constant density in the radial direction and constant
density in the polar direction ($p=1$, $q=0$); as $q$ is allowed to
increase the effect is weaker or even absent.

\item Although the silicate emission feature could be suppressed in
the clumpy models for certain parameters, the smooth models are able
to reproduce almost the same range of the silicate feature strength.
Our analysis shows that, overall, when considering characteristics of
the silicate feature, models with the three dust configurations
(smooth, two-phase, clumps-only) are not distinguishable.

\item Low density dust, smoothly distributed between the clumps in
the two-phase model, significantly contributes to the near-infrared
emission in type 1 view. This is the main difference with respect to
the clumps-only models that typically show a deficiency in this
range. This peculiar characteristic of the two-phase models might
represent a possible solution to a similar issue found when fitting
observed SED with currently available clumpy models from the
literature. 

\end{itemize}

This work will be extended and the parameter grid will be
progressively improved; models with different chemical composition of
the dust, different torus and clumps sizes and their spatial
distributions will be further explored. SEDs,  in the form of
\texttt{ascii} files are available on the following address:
\texttt{https://sites.google.com/site/skirtorus/}. Images, in the
form of \texttt{fits} files are available upon request.

%--------------------------------------------------------------------
\section*{Acknowledgments}
%--------------------------------------------------------------------

We thank the anonymous referee for useful comments and suggestions.
This work was supported by the European Commission (Erasmus Mundus
Action 2 partnership between the European Union and the Western
Balkans, http://www.basileus.ugent.be) and by the Ministry of
Education and Science of Serbia through the projects
`Astrophysical Spectroscopy of Extragalactic Objects' (146001) and
`Gravitation and the Large Scale Structure of the Universe'
(146003). The European Commission and EACEA are not responsible for
any use made of the information in this publication.

%\appendix

\bsp

\label{lastpage}

\end{document}